\documentclass[aps,prd,twocolumn,showpacs,preprintnumbers,nofootinbib,amsmath,amssymb]{revtex4}

\usepackage{graphicx}
\usepackage{dcolumn}
\usepackage{bm}
\usepackage{amsmath}
\usepackage{braket}
\usepackage[normalem]{ulem}
\usepackage[hidelinks]{hyperref}
\usepackage{epstopdf}
\usepackage{placeins}
\usepackage{multirow}
\usepackage{makecell}
\usepackage{cleveref}
\usepackage{xcolor}
\usepackage{soul}
\usepackage{enumitem}

\usepackage[font=small, labelfont=bf, textfont={small,it}]{caption}
\newcommand{\beq}{\begin{eqnarray}}
\newcommand{\eeq}{\end{eqnarray}}
\newcommand{\beqnn}{\begin{eqnarray*}}
\newcommand{\eeqnn}{\end{eqnarray*}}

\newcommand{\dof}{\mathrm{dof}}
\newcommand{\mc}{\mathrm{mc}}

\newcommand{\CP}{\mathrm{CP}}
\newcommand{\SU}{\mathrm{SU}}
\newcommand{\cool}{\mathrm{cool}}
\newcommand{\topo}{\mathrm{topo}}
\newcommand{\plaq}{\mathrm{plaq}}

\def\spose#1{\hbox to 0pt{#1\hss}}
\def\ltapprox{\mathrel{\spose{\lower 3pt\hbox{$\mathchar"218$}}
 \raise 2.0pt\hbox{$\mathchar"13C$}}}

\begin{document}

\title{Topological susceptibility of $2d~\mathrm{CP}^1$ or $\mathrm{O}(3)$ non-linear $\sigma$-model: is it divergent or not?}

\author{Claudio Bonanno}
\email{claudio.bonanno@fi.infn.it}
\affiliation{INFN Sezione di Firenze,\\ 
Via G.~Sansone 1, I-50019 Sesto Fiorentino, Firenze, Italy
}

\author{Massimo D'Elia}
\email{massimo.delia@unipi.it}
\affiliation{Universit\`a di Pisa and INFN Sezione di Pisa,\\ 
Largo B.~Pontecorvo 3, I-56127 Pisa, Italy
}

\author{Francesca Margari}
\email{f.margari@studenti.unipi.it}
\affiliation{Universit\`a di Pisa and INFN Sezione di Pisa,\\ 
Largo B.~Pontecorvo 3, I-56127 Pisa, Italy
}

\date{\today}

\begin{abstract}
The topological susceptibility of $2d$ $\mathrm{CP}^{N-1}$ models is expected, based on perturbative computations, to develop a divergence in the limit $N \to 2$, where these models reduce to the well-known \emph{non-linear $\mathrm{O}(3)$ $\sigma$-model}. The divergence is due to the dominance of instantons of arbitrarily small size and its detection by numerical lattice simulations is notoriously difficult, because it is logarithmic in the lattice spacing. We approach the problem from a different perspective, studying the behavior of the model when the volume  is fixed in dimensionless lattice units, where perturbative predictions are turned into more easily checkable behaviors. After testing this strategy for $N = 3$ and $4$, we apply it to $N = 2$, adopting at the same time a multicanonic algorithm to overcome the problem of rare topological fluctuations on asymptotically small lattices. Our final results fully confirm, by means of purely non-perturbative methods, the divergence of the topological susceptibility of the $2d$ $\mathrm{CP}^1$ model.
\end{abstract}


\maketitle

\section{Introduction}
\label{sec:intro}

The $2d$ $\CP^{N-1}$ models are quantum field theories that play an important role in the study of the non-perturbative properties of gauge theories, as they share many intriguing features with $4d$ Yang--Mills theories, such as confinement, the existence of a non-trivial topological structure and the related dependence on the topological parameter $\theta$.~\cite{DAdda:1978vbw,advanced_topics_QFT,Vicari:2008jw}. These theories are amenable to be treated exactly by analytic means in certain regimes, but have also been extensively explored by means of Monte Carlo (MC) simulations on the lattice, since they constitute the perfect theoretical laboratory to test new numerical methods in view of an application to the more complicated physical gauge theories.

At large-$N$, $\CP^{N-1}$ models admit a $1/N$ expansion which is similar 
to the 't Hooft large-$N$ expansion of QCD. These models, however, admit an analytic solution in this 
regime, and the large-$N$ limit of the vacuum energy $E(\theta)$ is well 
known both analytically and numerically~\cite{Witten:1980sp, Witten:1998uka, Rossi:1993nz, Rossi:2016uce, Campostrini:1991kv, Campostrini:1992ar, Campostrini:1992it, Campostrini1993, Bonati:2016tvi, Bonanno:2018xtd, Berni:2019bch}.
An important difference between $2d$ $\CP^{N-1}$ models and $4d$ $\SU(N)$ 
Yang--Mills theories emerges in the opposite, small-$N$ limit. 
Indeed, in the $N\to 2$ limit, where the theory becomes 
equivalent to the non-linear $\mathrm{O}(3)$ $\sigma$-model (which has been 
widely studied both analytically and numerically in the 
literature~\cite{Forster:1977jv,Berg:1981er,Haldane:1982rj, Haldane:1983ru,Bhanot:1984rx,Haldane:1985xx,Wiegmann:1985jt,Affleck:1987ch,Hasenfratz:1990zz,DElia:1995zja, DElia:1995wxi, Blatter:1995ik,Burkhalter:2001hu,Controzzi:2003wp,Alles:2007br,Nogradi:2012dj,Azcoiti:2012ws,Alles:2014tta,Bietenholz:2018agd, Bruckmann:2018usp, Sulejmanpasic:2018upi,Sulejmanpasic:2020lyq,Thomas:2021ewk,Marino:2022ykm}), a pathological behavior emerges,
which has no analogue in the Yang--Mills case,
where instead the approach from small to large $N$ is much 
smoother~\cite{Bonati:2016tvi,Bonanno:2020hht,Kitano:2020mfk, Kitano:2021jho}. 

The semi-classical picture predicts a divergence 
of the topological susceptibility $\chi$ for $N=2$, 
which survives the renormalization procedure. Various studies have 
already tried to check this prediction by lattice numerical 
simulations, and while there is a general consensus that the prediction
is verified, the issue is not completely settled. 
For example, while various works about the non-linear 
$\mathrm{O}(3)$ $\sigma$-model 
found numerical evidence supporting that $\chi$ is divergent in the 
continuum limit (see, e.g., Refs.~\cite{Bietenholz:2018agd, Thomas:2021ewk}), a recent investigation~\cite{Berni:2020ebn}
from some of the authors of the present paper, 
considering both direct simulations at $N = 2$ and the 
$N\to2$ limit of $\CP^{N-1}$ models, pointed out some difficulties in 
making a definite statement.

The main difficulty can be related to the fact that
the divergence is of ultraviolet (UV) origin, i.e., it is related to the 
presence of semiclassical solutions with non-zero 
topological charge (instantons) at arbitrarily small scales.
As a consequence, lattice studies need to check 
the emergence of a divergent behavior as the lattice spacing 
$a \to 0$: this task can be ambiguous, since the behavior could 
be barely distinguishable from a badly convergent behavior 
in a wide range of lattice spacings.
For instance, even for $N = 3$ the finiteness 
of $\chi$ could be definitely established only recently
(see, e.g., Ref.~\cite{Berni:2020ebn}, where 
two different strategies led consistent results).

The purpose of the present study is to develop a novel strategy,
in order to make the problem better defined and reach more definite
conclusions. In practice, we will approach the continuum limit
keeping the ratio between the UV and the infrared (IR) cutoffs fixed,
i.e., working at fixed volume in lattice units, and then considering 
the same procedure for different values of the dimensionless
volume, a strategy which resembles some aspects of the determination
of the step scaling beta-function on the lattice (see, e.g., Ref.~\cite{DelDebbio:2021ryq} for a recent review on the topic).
As we will discuss in more details in the following, within
this framework the original divergent behavior of the 
topological susceptibility is turned into a convergent 
(as opposed to vanishing) continuum limit,
which is much easier to check numerically, 
as indeed we will manage to do.

A drawback of this strategy is that one is forced to study 
volumes of arbitrarily small size in physical units, where 
topological fluctuations are extremely rare and a precise 
determination of the topological susceptibility could 
require an unfeasible statistics. This problem is easily 
solved by adopting a 
\emph{multicanonical algorithm}~\cite{Berg:1991cf}, which
has been recently employed to face the same issue in 
Refs.~\cite{Jahn:2018dke,Bonati:2017woi, Bonati:2018blm, Athenodorou:2022aay}. The general idea is to add a bias potential to the action, so that the probability 
of visiting suppressed topological sectors is enhanced;
the MC averages with respect to the original 
distribution are then obtained by means of an \emph{exact} standard reweighting 
procedure.

This paper is organized as follows. 
In Sec.~\ref{sec:model} we give a brief review  
about $2d$ $\CP^{N-1}$ models and their topological properties.
In Sec.~\ref{sec:setup} 
we describe our numerical setup and our strategy to compute the 
continuum limit of $\chi$ on asymptotically small lattices,
including a description of the adopted multicanonical algorithm. 
In Sec.~\ref{sec:results} we present and discuss our numerical results,
including also an application of the same method to $N=3$ and $4$, 
in order to check consistency with previous results in the literature for these models~\cite{Petcher198353, Campostrini:1992it,Lian:2006ky,Berni:2020ebn}.
Finally, in Sec.~\ref{sec:final}, we draw our conclusions.

\section{Continum theory}
\label{sec:model}

The Euclidean action of $2d$ $\CP^{N-1}$ models can be written in terms of a matter field $z(x)$, a complex $N$-component scalar field satisfying $\bar{z}(x) z(x) = 1$, and of an auxiliary non propagating $\mathrm{U}(1)$ gauge field $A_\mu$. In the presence of the topological term, the action reads 
\beq\label{eq:total_action}
S(\theta)=\int d^2x\left[ \frac{N}{g}\bar{D}_\mu \bar{z}(x) D_\mu z(x) -i \theta q(x) \right],
\eeq
where $g$ is the 't Hooft coupling, $D_\mu=\partial_\mu+iA_\mu$ is the $\mathrm{U}(1)$ covariant derivative and
\beq\label{eq:def_topocharge}
Q=\int d^2x \, q(x)= \frac{1}{2\pi} \epsilon_{\mu\nu} \int d^2x \, \partial_\mu A_\nu(x) \in \mathbb{Z}
\eeq
is the integer-valued topological charge. 

The $\theta$-dependent vacuum energy density, using the path-integral formulation of the theory, is given by
\beq
E(\theta)=-\frac{1}{V}\log \int [d\bar{z}][dz][dA]e^{-S(\theta)},
\eeq
where $V$ is the $2d$ space-time volume. Assuming that $E(\theta)$ is an analytic function of $\theta$ around $\theta=0$, one can Taylor expand it around this point; at leading order, one has~\cite{Vicari:2008jw,Bonati:2016tvi}:
\beq\label{eq:free_energy_expansion}
E(\theta)-E(0)=\frac{1}{2}\chi\theta^2 + O(\theta^4),
\eeq
where $\chi$ is the topological susceptibility 
\beq
\chi = \frac{1}{V}\braket{Q^2}\bigg\vert_{\theta=0}.
\eeq

To better understand the origin of the divergence of $\chi$ in the $N \to 2$ limit,
it is useful to recall that, in the semi-classical approximation, the path-integral is evaluated by integrating fluctuations around instanton solutions, and it is reduced to an ordinary integral of the instanton density. At leading order, the instanton density of $\CP^{N-1}$ models is given, as a function of the instanton size, by~\cite{Luscher:1981tq}:
\beq\label{eq:instanton_density}
d_I (\rho) \propto \rho^{N-3}.
\eeq
For $N=2$, $d_I (\rho) \sim 1/\rho$, i.e., it develops an UV divergence for $\rho \to 0$. This means that the divergence of $\chi$ in this case can be traced back to the proliferation of small-size instantons with vanishing size $\rho\to 0$, whose density grows proportionally to $1/\rho$.

\section{Numerical methods}\label{sec:setup}

In this section we discuss various aspects related to the discretization
of the models and of the observables, in particular those related to topology, and to the employed numerical strategies.

\subsection{Discretization details}
\label{subsec:Discretization_details}
We discretized space-time through a square lattice with $L^2$ sites and periodic boundary conditions, and the $\theta=0$ continuum action~\eqref{eq:total_action} through the tree-level Symanzik-improved lattice action~\cite{Campostrini:1992ar}:
\beq\label{eq:Symanzik_improved_lattice_action_cpn}
\begin{aligned}
S_L = &-2N\beta_L\sum_{x,\mu} \left \{ c_1 \Re\left[\bar{U}_\mu(x)\bar{z}(x+\hat{\mu})z(x)\right] \right. \\
&\left. + c_2 \Re\left[\bar{U}_\mu(x+\hat{\mu})\bar{U}_\mu(x)\bar{z}(x+2\hat{\mu})z(x)\right]
\right\},
\end{aligned}
\eeq
where $c_1=4/3$ and $c_2=-1/12$ are improvement coefficients, $\beta_L \equiv 1/g_L$ is the inverse bare coupling, $z(x)$ are the matter fields, satisfying $\bar{z}(x) z(x) = 1$, and $U_\mu(x)$ are $\mathrm{U}(1)$ gauge link variables. Symanzik improvement cancels out logarithmic corrections to the leading continuum scaling~\cite{improved_action}, improving convergence towards the continuum limit.

In this limit, approached taking $\beta_L \to \infty$, a vanishing lattice spacing $a\to0$ can be traded for a divergent lattice correlation length $\xi_L \equiv \xi/a \underset{a\to0}{\sim} 1/a$. In order to fix $a$, in this work we chose the second moment correlation length $\xi$, defined in the continuum theory as
\beq\label{eq:def_xi}
\xi^2 \equiv \frac{1}{\int G(x)d^2x}\int G(x) \frac{\vert x\vert^2}{4} d^2 x ,
\eeq
where $G(x)$ denotes the two-point connected correlation function of the projector $P_{ij}(x) \equiv z_i(x) \bar{z}_j(x)$: 
\beq\label{eq:projector_definition}
G(x) \equiv \braket{P_{ij}(x)P_{ij}(0)}-\frac{1}{N} .
\eeq
A lattice discretization of Eq.~\eqref{eq:def_xi} can be obtained from the Fourier transform $\tilde{G}_L(p)$ of $G_L(x)$, which is the lattice counterpart of Eq.~\eqref{eq:projector_definition}~\cite{Caracciolo:1998gga}:
\beq\label{eq:def_xi_L}
\xi_L^2 = \frac{1}{4\sin^2\left(\pi/L\right)}\left[ \frac{\tilde{G}_L(0,0)}{\tilde{G}_L(2\pi/L,0)}-1 \right].
\eeq

\subsection{Topology on the lattice and smoothing}

There are several possible discretizations $Q_L$ of the topological charge~\eqref{eq:def_topocharge}, 
all yielding the same continuum limit for the topological susceptibility and other quantities
relevant to $\theta$-dependence, when discretization effects are properly taken care of. Generally speaking, lattice definitions are related to the continuum one by~\cite{Campostrini:1988ab, Farchioni:1993jd}:
\beq\label{eq:lattice_charge}
Q_L = Z_Q(\beta_L) Q,
\eeq
where $Z_Q(\beta_L)$ is a finite multiplicative renormalization factor. For this reason, lattice discretizations of $Q$ are in general not integer-valued. The most simple discretization can be defined in terms of the plaquette $\Pi_{\mu \nu} (x) \equiv U_\mu(x) U_\nu(x+\hat{\mu})\bar{U}_\mu(x+\hat{\nu})\bar{U}_\nu(x)$ as:
\beq
Q_\plaq = \frac{1}{2\pi}\sum_{x} \Im \left[\Pi_{12}(x)\right].
\eeq

However, it is possible to work out \emph{geometric} discretizations of the topological charge~\cite{Berg:1981er, Campostrini:1992ar}, which always result in integer values for every configuration, i.e., definitions with $Z_Q=1$. In particular, we adopted the geometric definition that can be built from the link variables $U_{\mu}(x)$~\cite{Campostrini:1992ar}:
\beq\label{eq:topo_charge_geo}
Q_U= 
\frac{1}{2\pi}\sum_{x} \Im \left\{ \log \left[\Pi_{12}(x)\right] \right\} \in \mathbb{Z}.
\eeq

Although $Q_U$ has $Z_Q=1$, renormalization effects are still present when computing $\chi$ because of \emph{dislocations}~\cite{Berg:1981nw, Campostrini:1992it}. Dislocations are UV fluctuations of the background gauge field that make establishing the winding number of the configuration ambiguous. The net effect is that dislocations result in an additive renormalization when computing the lattice topological susceptibility~\cite{Alles:1997nu,DElia:2003zne}. Such renormalization diverges in the continuum limit and thus must be removed.

Being dislocations the result of UV fluctuations at the scale of the lattice spacing, computing the geometric charge on smoothed configurations is sufficient to remove their unphysical contribution, while preserving the background topological structure of the gauge fields. Indeed, smoothing brings a configuration closer to a local minimum of the action, thus dumping UV fluctuations while, at the same time, preserving the physical topological signal.

Many different smoothing algorithms have been proposed in the literature, such as stout smearing, gradient flow, or cooling, all giving consistent results when properly matched with each other (see Refs.~\cite{Bonati:2014tqa, Alexandrou:2015yba} for more details). For this reason, we chose cooling for its numerical cheapness. This method consists in a sequence of $n_\cool$ steps in which the configuration approaches a local minimum of the action by iteratively aligning both link variables $U_{\mu}(x)$ and site variables $z(x)$ to their relative local force. Since the choice of the action that is locally minimized during cooling is irrelevant~\cite{Alexandrou:2015yba}, we adopted the unimproved one for this purpose, meaning that the local forces along which the $U_\mu(x)$ and $z(x)$ fields are aligned are computed from the action
in Eq.~\eqref{eq:Symanzik_improved_lattice_action_cpn} with $c_1 = 1$ and $c_2 = 0$. In the end, thus, we define:
\beq
\begin{aligned}
Q_L &= Q_U^{(\cool)}, \\
a^2\chi &= \frac{\braket{Q_L^2}}{L^2}.
\label{eq:def_chi_lattice}
\end{aligned}
\eeq

It is worth mentioning that smoothing methods act as diffusive processes, thus modifying the UV behavior of the fields below a \emph{smoothing radius} $r_s$ which is proportional to the square root of the amount of smoothing performed (e.g., to $\sqrt{n_\cool}$ in our case). When $\chi$ is finite, 
the choice of $n_\cool$ is not critical because the physical topological 
signal is well separated from the length scale $r_s$ introduced by the 
smoothing procedure. As a consequence, in such cases $\chi$ exhibits a plateau upon increasing $n_\cool$ above a certain threshold, and no residual dependence on $n_\cool$ is observed on continuum-extrapolated results.

The pathological case of the $\CP^1$ model is, instead, different in this 
respect, since we exactly aim at probing the sensitivity of $\chi$ to the 
contribution of small instantons, which are however 
smoothed away below $r_s$ (see Ref.~\cite{Bietenholz:2018agd}, 
where the dependence of the topological susceptibility of the non-linear
$\mathrm{O}(3)$ $\sigma$ model on the gradient flow time is discussed).
In particular, in our setup where $L=l/a$ is kept fixed 
as $a \to 0$, the quantity $\sqrt{n_\cool}/L = r_s/l$ is a relevant parameter and we expect the continuum limit of $\chi$ to depend on its value. Thus, we will extrapolate our results towards $r_s/l\to0$ in order to ensure that no relevant contribution coming from small length scales is lost.

Concerning updating algorithms, $\CP^{N-1}$ models at small $N$
do not require any particular strategy to decorrelate the topological
charge~\cite{Berni:2020ebn}, contrary to the large-$N$ case, which
is plagued by a severe \emph{topological critical slowing down}~\cite{DelDebbio:2004xh, Luscher:2011kk, Hasenbusch:2017unr, Bonanno:2018xtd, Berni:2019bch, Bonanno:2020hht, Bonanno:2022yjr}. As a matter of fact, $\CP^{N-1}$ models at small $N$ are dominated by
small instantons, which are more easily decorrelated by means of local field 
updates. Thus, we will adopt local updating algorithms such as the Over-Relaxation (OR) and the over-Heat-Bath (HB)~\cite{Campostrini:1992ar}.
An issue is however represented by the dominance of the $Q = 0$ sector
on small physical volumes, which is discussed in more details in Sec.~\ref{subsec:dominance_of_Q_zero}.

\subsection{Continuum limit at fixed volume in lattice units}
\label{subsec:strategy}

The expectation value of a generic observable $\mathcal{O}$ scales towards the continuum limit according to 
\beq
\label{eq:UV_corrections}
\braket{\mathcal{O}}_L\left(\xi_L\right) = \braket{\mathcal{O}}_{\mathrm{cont}} + c\,\xi_L^{-2} + o\left(\xi_L^{-2}\right),
\eeq
where finite lattice spacing corrections to continuum scaling are expressed as inverse powers of $1/\xi_L$. However, the continuum scaling of topological observables is modified at small-$N$, due to the presence of small-size topological fluctuations. 

Such modifications have been worked out in Ref.~\cite{Berni:2020ebn} assuming the perturbative computation of the instanton size distribution $d_I(\rho) \propto \rho^{N-3}$ and that topological fluctuations are dominated by a non-interacting gas of small-size instantons and anti-instantons. Under these assumptions, the number of (anti-)instantons $n_I$ ($n_A$) is distributed as a Poissonian with $\braket{n_I} = \braket{n_A} \propto l^2 \int_{\rho_{\min}}^{\rho_{\max}} \rho^{N-3} d\rho$, where the integral is taken from a UV scale $\rho_{\min}$,
proportional to the lattice spacing $a$, to a IR scale 
$\rho_{\max}$, proportional to the correlation length $\xi$. Then,
\beq
\braket{Q^2} \propto \braket{ \left( n_I-n_A \right)^2 } = 2 \braket{n_I} \propto l^2 \int_{\rho_{\min}}^{\rho_{\max}} \rho^{N-3} d\rho,
\eeq
and, thus,
\beq \label{eq:chi_small_N_continuum_scaling}
\xi^2\chi = \xi^2 \frac{\braket{Q^2}}{l^2}  \propto
\begin{cases}
\rho_{\max}^{N-2}-\rho_{\min}^{N-2}, &(N>2),
\\
\\
\log\left(\dfrac{\rho_{\max}}{\rho_{\min}}\right), &(N=2).
\end{cases}
\eeq
From Eq.~\eqref{eq:chi_small_N_continuum_scaling}, taking
into account that $\rho_{\min} \propto a$ and $\rho_{\max} \propto \xi$,
one can predict the following behaviors for the topological susceptibility when the continuum limit is approached at fixed physical lattice 
volume $V=l^2$
(hence, on lattices satisfying $L/\xi_L \gg 1$):
\begin{align*}
\xi^2\chi(x) &= A_2 \log(B_2 x) + C_2 x^2 + O(x^4), &\,\, (N=2),\\
\xi^2\chi(x) &= A_3 + B_3 x + C_3 x^2 + O(x^4) , &\,\, (N=3),\\
\xi^2\chi(x) &= A_N + C_N x^2 + O(x^4), &\,\, (N \ge 4),
\end{align*}
where $x = 1/\xi_L \propto a$. Hence, for the $\CP^1$ model the divergence of the topological susceptibility should appear as a logarithm of the UV cut-off
$1/a$, which may be difficult to distinguish from a regular power-law behavior in $a$. 

To overcome this issue, we investigate the continuum limit of $\chi$ in the small-$N$ limit performing lattice simulations at fixed volume in lattice units, i.e., fixing $L=l/a$.
Using this approach, we have the following predictions:
\beq
\label{eq:small_N_UV_corrections}
\xi^2\chi = \xi^2 \frac{\braket{Q^2}}{l^2} \propto
\begin{cases}
a^{N-2}\left[\left(\dfrac{L}{R}\right)^{N-2} - 1 \right]  &(N>2),
\\
\\
\log\left(\dfrac{L}{R}\right) &(N=2),
\end{cases}
\eeq
where $R$ is an effective parameter accounting for the 
ratio between the maximum and the minimum instanton sizes 
which can live on the same lattice, which is expected in this case to be 
proportional to $l = a L$ (since $L \ll \xi_L$), i.e., 
$\rho_{\max}/\rho_{\min} \equiv L/R$, with $R$ independent of $L$. 
We stress that Eq.~\eqref{eq:small_N_UV_corrections} has been obtained by multiplying $\braket{Q^2}/l^2$ for the squared correlation length $\xi^2$ obtained on large physical volumes (i.e., in the limit $L/\xi_L \gg 1$), so that the only dependence on $L$ of the continuum limit of $\xi^2 \chi$ comes from $\chi$ alone. 

These semi-classical considerations point out that, when the continuum topological susceptibility computed on lattices with $L/\xi_L \gg 1$ is finite, the continuum limit of $\chi$ at fixed $L$ is expected to vanish as $a$ for $N=3$ or as $a^2$ for $N=4$, cf.~Eq.~\eqref{eq:small_N_UV_corrections}. This is due to the fact that, when $a \to 0$ at fixed $L$, the physical lattice size vanishes proportionally to $a$, and any topological fluctuation on physical scales disappears.

On the other hand, if the continuum limit of $\chi$ taken at fixed $L/\xi_L \gg 1$ is logarithmically divergent, as predicted by semi-classical computations, 
we expect to approach a constant and finite value for $\chi$ 
when, instead, the continuum limit is taken at fixed $L$, 
cf.~again Eq.~\eqref{eq:small_N_UV_corrections}. In this case, topological fluctuations damped because of the decreasing IR cut-off are exactly balanced, as $a \to 0$, by new topological fluctuations appearing at arbitrarily small UV scales.
This means that, with this strategy, the divergent continuum limit of $\chi$
predicted by semiclassical computations is mapped into a non-vanishing finite 
continuum limit, which should be more amenable to be tested by 
numerical methods.

\subsection{Determining $\xi_L$ close to the continuum limit}
\label{subsec:determining_xiL}

As we stressed above, the values of $\xi_L$ needed for our determination
of $\xi^2 \chi = \xi_L^2 a^2 \chi$ are those that would be 
obtained in the infinite volume limit, i.e., on lattices of size $L$ 
such that $L/\xi_L \gg 1$.
This is barely feasible within the framework of our numerical strategy,
aimed at reaching very large values of $\xi_L$ but keeping $L$ fixed, 
i.e.~we would need to perform additional simulations on unfeasible large lattices
for a reliable numerical determination of $\xi_L$.

To overcome this problem, we will first look for the onset
of the asymptotic scaling region where $\xi_L$ scales as predicted 
by the $2$-loop perturbative beta-function~\cite{advanced_topics_QFT}
\beq\label{eq:2loop_beta_function}
-a \frac{d \beta_L^{-1}}{da} = -\frac{\beta_L^{-2}}{2\pi}\left(1+\frac{\beta_L^{-1}}{\pi N}\right) \, , 
\eeq
and then make use of such scaling to extend the determination of $\xi_L$
within this region.

Integrating Eq.~\eqref{eq:2loop_beta_function} to obtain the running of the quantity $2\pi\beta_L(a)$, it is possible to obtain the dynamically-generated scale of the lattice theory~\eqref{eq:Symanzik_improved_lattice_action_cpn} $\Lambda_L^{(\mathrm{Sym})}$ in lattice units~\cite{Campostrini:1992ar}
\beq
\Lambda_L^{(\mathrm{Sym})}  = \frac{1}{a} [(2 \pi \beta_L)^{2/N} \exp\{-2\pi\beta_L\}] \equiv \frac{1}{a} f(\beta_L).
\eeq
The latter equation can be turned into a perturbative expression for the mass gap $M \equiv \xi^{-1}$ by multiplying both sides for $\xi$:
\beq\label{eq:mass_gap_2loop}
\Lambda^{\mathrm{(Sym)}}_L / M = \xi_L f(\beta_L).
\eeq
Being Eq.~\eqref{eq:mass_gap_2loop} the result of a perturbative computation, we expect the ratio $M/\Lambda^{\mathrm{(Sym)}}_L$ to approach a constant value plus $O(1/\beta_L)$ corrections in the asymptotic region $\beta_L\to\infty$. Assuming such corrections to be negligible, Eq.~\eqref{eq:mass_gap_2loop} allows to compute $\xi_L$ at arbitrarily-large values of the bare coupling once its value $\xi_L^\star$ for a certain coupling $\beta_L^\star$ is fixed:
\beq\label{eq:xi_2loop}
\xi_L(\beta_L) = f(\beta_L^\star) \frac{\xi_L^\star}{f(\beta_L)}.
\eeq
In the following, $\xi_L$ will be first determined numerically 
on large lattices (satisfying $L/\xi_L \gg 1$) and for a feasible range of values of
$\beta_L$; then, by matching results to asymptotic scaling 
prediction in Eq.~\eqref{eq:mass_gap_2loop}, we will choose a $\beta_L^\star$ for which
Eq.~(\ref{eq:xi_2loop}) is reliable, and determine 
$\xi_L$ accordingly for larger values of $\beta_L$. More details about our choice of $\beta_L^{\star}$ and on the check of the stability of $\xi_L$ varying this choice can be found in App.~\ref{app:asymptotic_scaling}.

\subsection{Dominance of the $Q=0$ sector and multicanonical algorithm}
\label{subsec:dominance_of_Q_zero}

Another drawback of working at fixed $L$ is the dominance of the $Q=0$ sector, which introduces the necessity of collecting unfeasible  statistics to achieve a precise computation of the topological susceptibility on asymptotically small lattice volumes. In order to better clarify this statement we remark that, 
according to Eq.~\eqref{eq:small_N_UV_corrections}, even if $\chi$ 
diverges for $N = 2$ as expected from the semi-classical approximation, 
$\braket{Q^2}$ is expected to vanish, for fixed $L$, 
as $1 / \xi_L^2$ in the continuum limit. If $\braket{Q^2} \ll 1$ then $P(Q = 0) \gg P(\vert Q\vert = 1) \gg P(\vert Q \vert = 2) \gg \dots$; in this regime, the variance of the topological charge distribution $P(Q)$ can be approximated as
\beq\label{eq:P_1_over_P_0}
\braket{Q^2} = V \chi \simeq \frac{P(\vert Q \vert=1)}{P(Q=0)},
\eeq
i.e., to compute $\chi$ we need to estimate with great precision a vanishing 
probability to visit $\vert Q \vert = 1$ sectors. This requires a growing and unfeasible numerical effort, since we need a sufficient number of fluctuations of $Q$ to obtain $\braket{Q^2}$ with a given target precision.

In order to overcome this problem we will adopt the multicanonical algorithm. This approach was recently employed in the context of $4d$ gauge theories to enhance topological fluctuations at finite temperature, see, e.g., Refs.~\cite{Jahn:2018dke,Bonati:2018blm}. The main idea behind the multicanonic approach is to modify the probability distribution of the topological charge $P(Q) \to P_{\mc}(Q) = P(Q)w(Q)$, where $w(Q)$ is a known $Q$-dependent weight function, in order to enhance the probability of visiting suppressed topological sectors. Since the relative error on~\eqref{eq:P_1_over_P_0} scales as the inverse of the square root of the number of $\vert Q \vert = 1$ events $\sim N_{\mathrm{meas}}\, P(\vert Q \vert=1)$, enhancing $P(\vert Q \vert=1)$ with respect to $P(Q=0)$ by a known factor of $w_1/w_0$ reduces the relative error on $\chi$ by a factor of $\sqrt{w_1/w_0}$.

In analogy with lattice QCD simulations~\cite{Bonati:2017woi,Jahn:2018dke,Bonati:2018blm}, we introduce the weights $w(Q)$ by adding a topological potential $V_{\topo}(Q_{\mc})$ to the lattice action:
\beq
\label{eq:multicano_action}
S_L \to S_L + V_{\topo}(Q_{\mc}) \implies w(Q) = e^{-V_{\topo}(Q_{\mc})},
\eeq
where $Q_{\mc}$ is a suitable discretization of the topological charge, which does not necessarily need to coincide with the one that is used to measure it. 

Expectation values with respect to the original distribution are then \emph{exactly} recovered by the following standard reweighting procedure:
\beq
\label{eq:reweighting}
\braket{O} = \frac{\braket{O e^{V_{\topo}(Q_{\mc})}}_{\mc}}{\braket{ e^{V_{\topo}(Q_{\mc})}}_{\mc}}.
\eeq

We stress that the relation in Eq.~\eqref{eq:reweighting} among expectation values computed with and without the bias potential is exact, thus, any choice for $V_{\topo}(Q_\mc)$ will in the end give the correct result for $\braket{O}$. Therefore, this strategy does not introduce any further source of uncertainty. For this reason, the discretization of the topological charge $Q_{\mc}$ and the bias potential $V_{\topo}$ can be chosen with some arbitrariness.

However, the choice of $V_{\topo}(Q_\mc)$ can affect the efficiency of the algorithm, and in particular one would like to avoid possible issues due to a poor overlap between the starting and the biased path-integral distributions. This could happen, e.g., if $V_\topo$ is too strong. In that case, a sort of spontaneous breaking of the $\CP$ symmetry occurs~\cite{Bonati:2018blm}, meaning that configurations with $Q \neq 0$ occur with overwhelming frequency with respect to $Q = 0$ ones and that $\braket{Q} \neq 0$, thus disrupting importance sampling and leading to uncontrolled effects on the correct estimation of statistical errors in the evaluation of the ratio of expectation values in Eq.~\eqref{eq:reweighting}.

However, such pathological cases can be easily avoided by tuning the topological potential through short test runs, ensuring that importance sampling is not disrupted. In particular, if the MC evolution of the topological charge is still dominated by the $Q = 0$ sector, the symmetry properties of the distribution are preserved (i.e., $\braket{Q} = 0$), and $Q \ne 0$ sectors (which are those giving contribution to the averages of interest) are explored more frequently, and then reweighted, that enhances (rather than disrupting) importance sampling for the observables of interest. The estimate of the statistical error on Eq.~\eqref{eq:reweighting}, which proceeds usually through a bootstrap analysis, will then be reliable if the number of tunnelings in and out of the $Q = 0$ topological sector is statistically significant, as is always the case in our simulations. The tuning of the potential was done following the procedure outlined in Ref.~\cite{Bonati:2018blm}. More details about our choices for $Q_{\mc}$ and $V_{\topo}$ and about our implementation of the multicanonical algorithm can be found in App.~\ref{app:details_multicano_algorithm}.

\section{Numerical results}
\label{sec:results}

In this section, we will first discuss results for the topological susceptibility for $N=4$ and $N=3$, showing that our strategy gives compatible results with previous findings in the literature~\cite{Petcher198353, Campostrini:1992it, Lian:2006ky,Berni:2020ebn}. Then, we show the behavior of $\xi^2 \chi$ in the continuum limit at fixed $L$ for $N=2$, established adopting the multicanonic algorithm. Finally, we conclude our study by comparing results achieved at fixed $L$ with those obtained at fixed physical volume ($L/\xi_L \gg 1$).

\subsection{Results for $N = 4$ and $N=3$}
\label{subsec:results_N_3_and_4}

In order to calibrate our strategy, we first consider the cases $N=4$ and $3$, for which we expect from semi-classical computations, and we actually know from previous lattice results~\cite{Petcher198353, Campostrini:1992it, Lian:2006ky,Berni:2020ebn}, that the topological susceptibility is finite.

Therefore, according to Eq.~\eqref{eq:small_N_UV_corrections}, we expect to observe a vanishing continuum limit
\beq\label{eq:semiclassical_N_4_and_3}
\xi^2 \chi(x) \underset{x\to 0}{\sim} x^c, \quad x = 1/\xi_L,
\eeq
where $c=2$ for $N=4$ and $c=1$ for $N=3$.

Following the strategy discussed in Sec.~\ref{subsec:strategy}, we performed lattice simulations keeping the volume fixed in lattice units on lattices with $L=50$, exploring several values of $\beta_L$ and reaching values of $\xi_L$ of the order of $\sim 10^3$. Our MC updating step in this case consisted of 4 lattice sweep of OR and 1 lattice sweep of HB updating steps: in the following we will simply call this combination ``standard MC step''. The computation of the topological susceptibility in lattice units via Eq.~\eqref{eq:def_chi_lattice} was performed every 10 MC steps and after $n_\cool=50$ cooling steps, while $\xi_L$ was computed via Eq.~\eqref{eq:xi_2loop}.

In Tab.~\ref{tab:summary_results_N_4_and_3} we report a complete summary of the parameters of the performed simulations for $N=4$ and $3$ along with the generated statistics and the obtained results for $\xi_L$, $a^2 \chi$ and $\xi^2 \chi$.

\begin{table}[!t]
\begin{center}
\begin{tabular}{|c|c|c|c|c|c|}
\hline
$N$ & $\beta_L$ & $\xi_L \cdot 10^{-3}$ &  $a^2 \chi\cdot 10^{9}$ & $\xi^2 \chi\cdot 10^{3}$ & Stat.\\
\hline
\multirow{9}{*}{$4$}
& 1.35 & 0.08262(70) & 283.0(2.7) & 1.932(38) & \multirow{9}{*}{52M} \\
& 1.40 & 0.11108(94) & 89.3(1.5)  & 1.102(27) & \\
& 1.45 & 0.1494(13)  & 28.55(87)  & 0.638(22) & \\
& 1.50 & 0.2012(17)  & 7.80(43)   & 0.316(18) & \\
& 1.55 & 0.2709(23)  & 2.88(27)   & 0.211(20) & \\
& 1.60 & 0.3651(31)  & 0.98(17)   & 0.130(22) & \\
& 1.65 & 0.4922(42)  & 0.42(11)   & 0.102(27) & \\
& 1.70 & 0.6639(56)  & 0.107(54)  & 0.047(24) & \\
& 1.75 & 0.8959(76)  & 0.031(19)  & 0.025(15) & \\
\hline
\multirow{14}{*}{$3$}
& 1.50  & 0.12765(68) & 529.1(4.1) & 8.62(11) & \multirow{6}{*}{25M} \\
& 1.55  & 0.17099(92) & 223.8(2.7) & 6.54(10) & \\
& 1.60  & 0.2292(12)  & 90.8(1.7)  & 4.77(10) & \\
& 1.65  & 0.3074(16)  & 38.4(1.1)  & 3.62(11) & \\
& 1.70  & 0.4126(22)  & 15.66(73)  & 2.67(13) & \\ 
& 1.75  & 0.5541(30)  & 6.87(47)   & 2.11(15) & \\ \cline{6-6}
& 1.80  & 0.7445(40)  & 2.78(15)   & 1.54(8)  & \multirow{2}{*}{102M} \\
& 1.85  & 1.0009(54)  & 1.18(10)   & 1.18(10) & \\ \cline{6-6}
& 1.90  & 1.3461(72)  & 0.501(76)  & 0.91(14) & \multirow{3}{*}{76M} \\
& 1.95  & 1.8114(97)  & 0.146(37)  & 0.48(12) & \\
& 2.00  & 2.438(13)	  & 0.083(40)  & 0.50(24) & \\ \cline{6-6}
& 2.05  & 3.284(18)	  & 0.025(11)  & 0.27(12) & \multirow{3}{*}{128M} \\
& 2.10  & 4.424(24)	  & 0.0156(83) & 0.31(16) & \\
& 2.15  & 5.963(32)	  & 0.0094(70) & 0.33(25) & \\
\hline
\end{tabular}
\end{center}
\caption{Summary of simulation parameters and results obtained for $N=4,3$ and $L=50$. The correlation length $\xi_L$ is computed according to Eq.~\eqref{eq:xi_2loop} with $\beta_L^\star = 1.35$ and $1.455$ for $N=4,3$ respectively (see App.~\ref{app:asymptotic_scaling} for more details). Reported values of $\chi$ are computed after $n_\cool = 50$ cooling steps. Statistics is expressed in millions (M) and measures are taken every 10 standard MC steps ($=$ 4 OR + 1 HB lattice updating sweeps).}
\label{tab:summary_results_N_4_and_3}
\end{table}

We start our discussion from the $\CP^3$ model. We extrapolated the quantity $\xi^2 \chi$ towards the continuum limit fitting the $\xi_L$-dependence of $\xi^2 \chi$ according to the fit function
\beq\label{eq:ansatz_cont_limit}
f(x) = a_0 + a_1 \, x^c, \quad\qquad x=1/\xi_L,
\eeq
where $c$ is a free exponent.

\begin{figure}[!htb]
\centering
\includegraphics[scale=0.35]{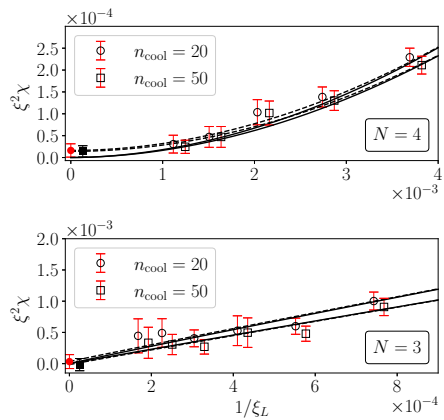}	
\caption{Continuum extrapolation of $\xi^2 \chi$ for $N=4$ (top) and $N=3$ (bottom). Solid and the dashed lines represent, respectively, best fits obtained using fit function~\eqref{eq:ansatz_cont_limit} setting $a_0=0$ and treating it as a free parameter. Determinations obtained for different values of $n_\cool$ have been slightly shifted to improve readability. Full points in $1/\xi_L = 0$ represent continuum-extrapolated determinations.}
\label{fig:fit_continuum_limit_N_4_3}
\end{figure}

In order to check that the continuum limit is indeed vanishing, we considered two cases: the case when $a_0$ is treated as a free parameter and the one when $a_0$ is fixed to zero. In the former case, our data turn out to be well compatible with a vanishing continuum limit, as the best fit yields $\tilde{\chi}^2/\dof = 6.6/6$ and $a_0=1.4(1.3)\cdot 10^{-5}$, which is compatible with zero within its statistical error. Moreover, the exponent $c=1.96(6)$ turns out to be compatible with 2, which is in agreement with the semi-classical prediction in Eq.~\eqref{eq:semiclassical_N_4_and_3} and with results of Ref.~\cite{Berni:2020ebn}. Also fixing $a_0=0$ gives a very good description of our data, as the best fit gives $\tilde{\chi}^2/\dof = 7.8/7$ and a compatible exponent $c=1.91(4)$. In Fig.~\ref{fig:fit_continuum_limit_N_4_3} we show such continuum extrapolations for $N=4$ considering all available determinations in Tab.~\ref{tab:summary_results_N_4_and_3}.

Varying the fit range, the value of $n_\cool$ used to compute $Q_L$ or the coupling $\beta^\star_L$ used to fix $\xi_L$ did not result in any appreciable variation of our final results. We can thus conclude that our findings are perfectly compatible with previous results in the literature pointing out a finite continuum limit for $\chi(N=4)$ (see, e.g., Refs.~\cite{Campostrini:1992it, Lian:2006ky, Berni:2020ebn}).

We now repeat the same analysis for $N = 3$. The continuum extrapolation of $\xi^2\chi$ data for $N = 3$ according to fit function~\eqref{eq:ansatz_cont_limit} is depicted in Fig.~\ref{fig:fit_continuum_limit_N_4_3}. The best fit in the whole available range yields $a_0 = (-2 \pm 10) \cdot 10^{-5}$, $c=0.98(4)$ and $\tilde{\chi}^2/\dof = 4.4/11$. Also performing the best fit fixing $a_0=0$ perfectly describes our data, giving $c=0.99(2)$ and $\tilde{\chi}^2/\dof=4.4/12$.

Again, we observe no dependence of our continuum-extrapolated results on the choice of $n_\cool$, of $\beta_L^\star$ or of the fit range. Therefore, also in this case our strategy gives compatible results both with semi-classical expectations and with previous numerical results in the literature for $\chi(N=3)$ (see, e.g., Refs.~\cite{Petcher198353, Berni:2020ebn}).

\subsection{Results for the topological susceptibility of the $\CP^1$ model from the multicanonic algorithm}
\label{subsec:multicano_results}

\begin{figure}[!htb]
\centering
\includegraphics[scale=0.35]{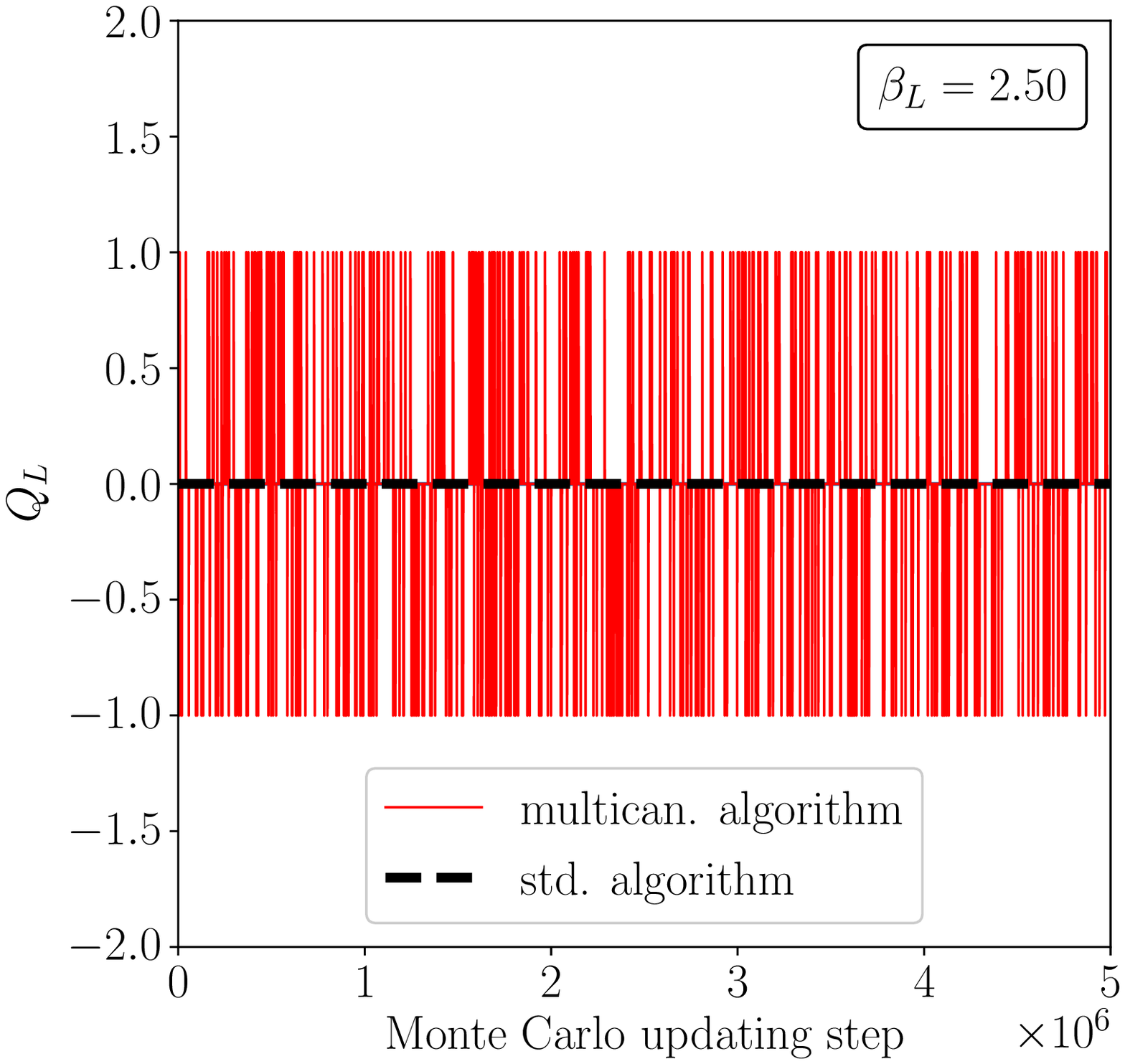}
\includegraphics[scale=0.35]{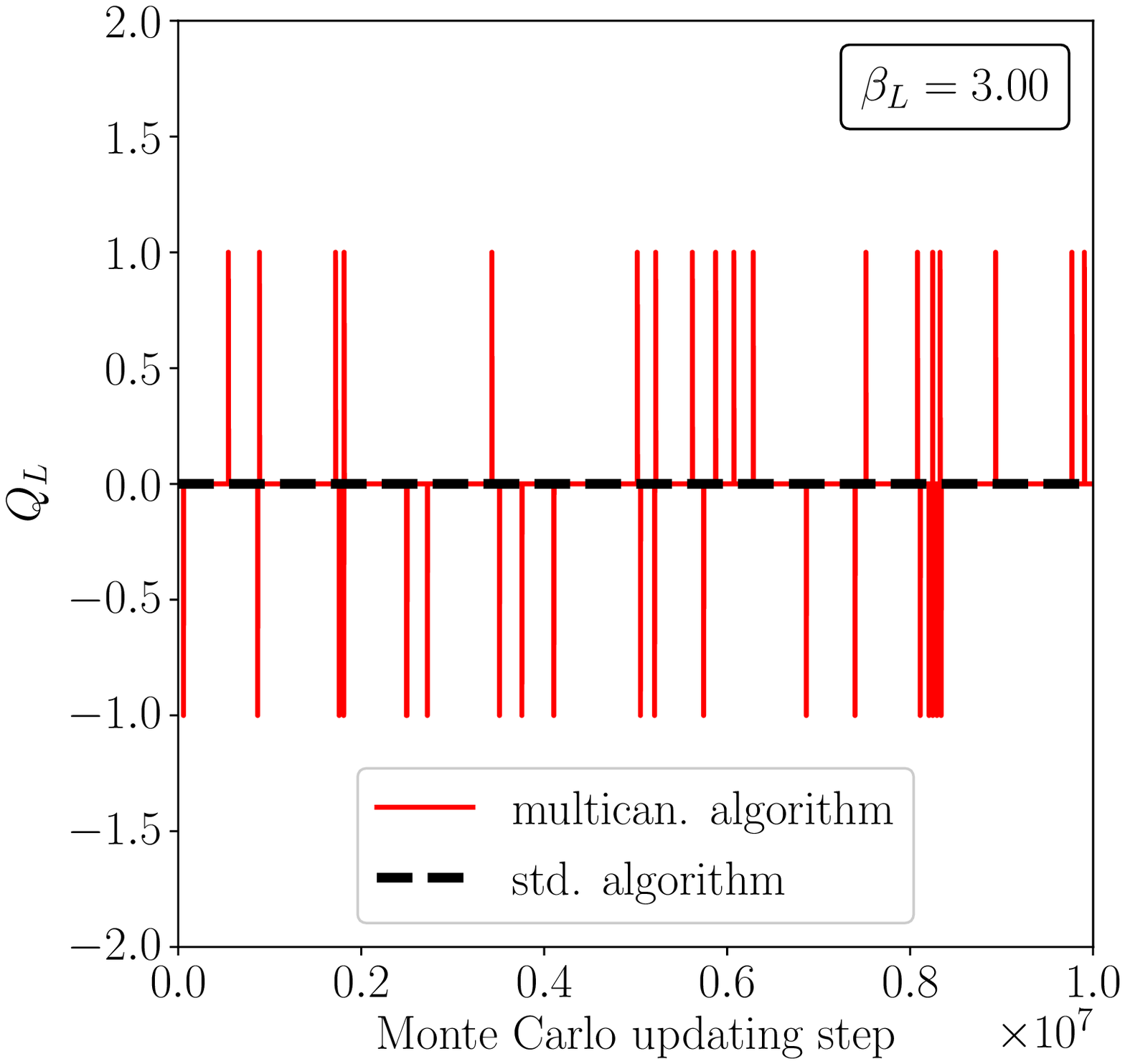}
\caption{Evolution of the geometric charge $Q_L$ computed after $n_\cool=50$ cooling steps for $N = 2$ obtained with the standard and the multicanonic algorithm.}
\label{fig:Q_U_cool_comparison}
\end{figure}

In order to precisely assess the continuum behavior of $\xi^2 \chi(N=2)$, we pushed our investigation on the $L=50$ lattice up to $\xi_L$ as large as $\sim 10^6$. Reliably computing the susceptibility for such fine lattice spacings is an unfeasible task with standard methods due to the dominance of the $Q=0$ sector previously explained, while it was made possible by the adoption of the multicanonic algorithm, which allowed to largely improve the number of fluctuations of $Q_L$ observed during MC simulations. Illustrative examples for $\beta_L=2.50$ and $3.00$ are shown in Fig.~\ref{fig:Q_U_cool_comparison}.

\begin{figure}[!htb]
\centering
\includegraphics[scale=0.35]{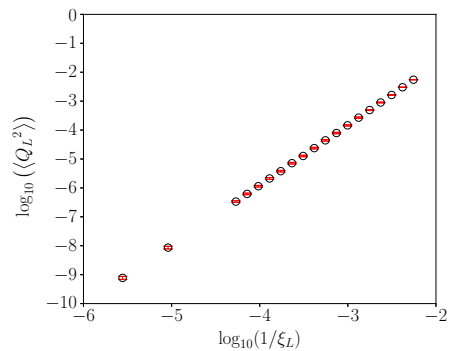}	
\caption{Behavior of $\braket{Q_L^2}$, measured after $n_\cool=50$ cooling steps, as a function of $1/\xi_L$ for $N=2$ and $L=50$.}
\label{fig:log_Q2}
\end{figure}

This has in turn allowed to largely reduce the computational power needed to determine $\chi$ with a given precision. As an example, let as consider our largest $\beta_L$, for which $\braket{Q^2} \sim 10^{-9}$ (cf.~Fig.~\ref{fig:log_Q2}). Using Eq.~\eqref{eq:P_1_over_P_0} and assuming that the error on $\chi$ scales as the inverse of the square root of $N_{\mathrm{meas}} \, P(\vert Q \vert = 1)$, we can estimate that, to reach the same $\sim 2\%$ relative error on the susceptibility achieved with the multicanonical algorithm, with the standard algorithm we would have needed a statistics larger by about a factor of $\sim 100$, i.e., we gained two orders of magnitude in terms of computational power.

For this reason, we adopted the multicanonic algorithm for $\beta_L \ge 2.20$, i.e., for $\xi_L\gtrsim 3 \cdot 10^3$, where $\braket{Q^2} \lesssim 10^{-5}$. A summary of all the performed simulation and the obtained results for $N=2$ is reported in Tab.~\ref{tab:summary_multican_results_N_2}.
\begin{table}[!htb]
\begin{center}
\begin{tabular}{|c|c|c|c|c|c|c|}
\hline
$N$ & Alg. & $\beta_L$ & $\xi_L \cdot 10^{-3}$ &  $a^2 \chi\cdot 10^{9}$ & $\xi^2 \chi\cdot 10^{3}$ & Stat.\\
\hline
\multirow{19}{*}{$2$} &
\multirow{10}{*}{Std}
& 1.70  & 0.17991(78) & 2207.4(4.8) & 71.45(64) & \multirow{6}{*}{51M} \\
&& 1.75  & 0.2393(10)  & 1205.7(3.6) & 69.03(63) & \\
&& 1.80  & 0.3185(14)  & 652.6(2.7)  & 66.21(64) & \\
&& 1.85  & 0.4243(18)  & 359.0(2.0)  & 64.62(66) & \\
&& 1.90  & 0.5656(25)  & 196.7(1.5)  & 62.92(72) & \\	
&& 1.95  & 0.7545(33)  & 107.6(1.1)  & 61.24(82) & \\ \cline{7-7}
&& 2.00  & 1.0072(44)  & 58.15(57)   & 58.99(77) & 102M \\
&& 2.05  & 1.3453(58)  & 31.59(34)   & 57.17(79) & 153M \\
&& 2.10  & 1.7980(78)  & 17.63(20)   & 57.00(82) & 256M \\
&& 2.15  & 2.404(10)   & 9.50(12)    & 54.90(83) & 410M \\ \cline{2-7}
& \multirow{9}{*}{\makecell{Multi-\\can.}}
& 2.20 & 3.217(14)   & 5.086(59)     & 52.64(76) & 81M  \\
&& 2.25 & 4.307(19)   & 2.864(35)     & 53.12(79) & 133M \\
&& 2.30 & 5.768(25)   & 1.516(18)     & 50.43(75) & 266M \\
&& 2.35 & 7.729(34)   & 0.849(11)     & 50.73(79) & 595M \\
&& 2.40 & 10.362(45)  & 0.4576(56)    & 49.13(74) & 566M \\
&& 2.45 & 13.897(60)  & 0.2475(30)    & 47.80(71) & 640M \\
&& 2.50 & 18.645(81)  & 0.1346(16)    & 46.78(70) & 1G   \\
&& 2.80 & 109.643(48) & 0.003448(68)  & 41.45(89) & 7G   \\ 
&& 3.00 & 359.6(1.6)  & 0.0003093(69) & 39.98(96) & 16G   \\
\hline
\end{tabular}
\end{center}
\caption{Summary of simulation parameters and results obtained for $N=2$ and $L=50$. The correlation length $\xi_L$ is computed according to Eq.~\eqref{eq:xi_2loop} with $\beta_L^\star = 1.70$ (see App.~\ref{app:asymptotic_scaling} for more details). Reported values of $\chi$ are computed after $n_\cool = 50$ cooling steps. Statistics is expressed in millions/billions (M/G) and measures are taken every 10 standard MC steps ($=$ 4 OR + 1 HB lattice updating sweeps) or every 10 multicanonic steps (see App.~\ref{app:details_multicano_algorithm} for more details).}
\label{tab:summary_multican_results_N_2}
\end{table}

To extrapolate our finite-$\xi_L$ determinations towards the continuum limit, we consider again the fit function ansatz in Eq.~\eqref{eq:ansatz_cont_limit}, and we perform a best fit of all available data for $\xi^2 \chi$ as a function of $1/\xi_L$, both considering fixed $a_0=0$ and $a_0$ as a free parameter.

\begin{figure}[htb!]
\centering
\includegraphics[scale=0.45]{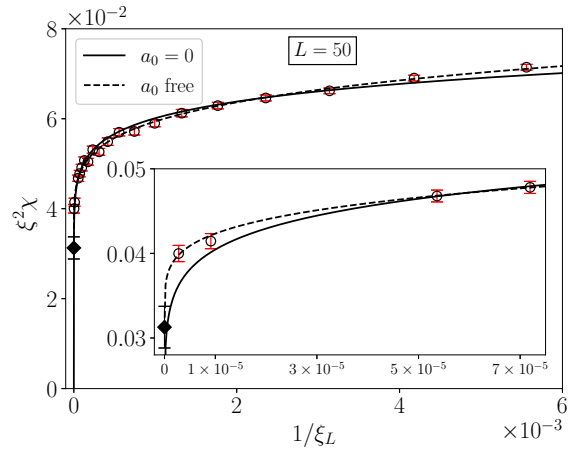}
\caption{Continuum extrapolation of $\xi^2 \chi$ for $N=2$, $L=50$ and $n_\cool=50$. Solid and the dashed lines represent, respectively, best fits obtained using fit function~\eqref{eq:ansatz_cont_limit} setting $a_0=0$ and treating it as a free parameter. The full point in $1/\xi_L = 0$ represents the non-vanishing continuum -extrapolated determination.}
\label{fig:continuum_limit_N_2_chi_multican}
\end{figure}

While in the latter case such best fit provides a very good description of our numerical results, giving 
\beqnn
a_0 &=& 0.031(2), \\
c &=& 0.20(2), \\
\tilde{\chi}^2/\dof &=& 8.0/16 ,
\eeqnn
the best fit performed fixing $a_0=0$ yields a $\tilde{\chi}^2/\dof = 42.7/17$, thus clearly providing a bad description of our data. Narrowing the fit range by, e.g., excluding the point at the smallest value of $\xi_L$ ($\beta_L=1.70$), does not improve the result, as we still obtain $\tilde{\chi}^2/\dof = 32.6/16$. On the other hand, the quality of the fit with $a_0$ free remains very good, as excluding the point for our smallest $\xi_L$ yields $a_0=0.031(3)$ with $\tilde{\chi}^2/\dof=7.5/15$. A comparison between the continuum limits taken at fixed $L$ in the whole available range is displayed in Fig.~\ref{fig:continuum_limit_N_2_chi_multican}.

It is interesting to observe that the best fit with $a_0$ free yield $\tilde{\chi}^2/\dof \simeq 0.5$, i.e., smaller than 1. A possible explanation is that, being all values of $\xi_L$ obtained from the same $2$-loop scaling equation, results for $\xi^2\chi$ at different values of $\beta_L$ are slightly correlated, thus our result for the $\tilde{\chi}^2/\dof$ is actually underestimated.
	
To check if this explanation is reasonable, we repeated the best fits previously discussed computing the error on $\xi^2\chi$ without considering the error on $\xi_L$, so that the mentioned correlation becomes irrelevant in the evaluation of the $\tilde{\chi}^2/\dof$. Treating $a_0$ as a free parameter, we obtain $a_0 = 0.033(2)$ with $\tilde{\chi^2}/\dof=17/16$, i.e., a perfectly agreeing result but with a $\mathcal{O}(1)$ reduced chi squared. The best fit with fixed $a_0=0$ is instead further disproved, as it yields an even larger reduced chi squared: $\tilde{\chi^2}/\dof=104/17$.

Summarizing, these results point out that $\xi^2 \chi$ behaves in the continuum limit in perfect agreement with the semi-classical prediction, cf.~Eq.~\eqref{eq:small_N_UV_corrections}.

\begin{figure}[!htb]
\centering
\includegraphics[scale=0.42]{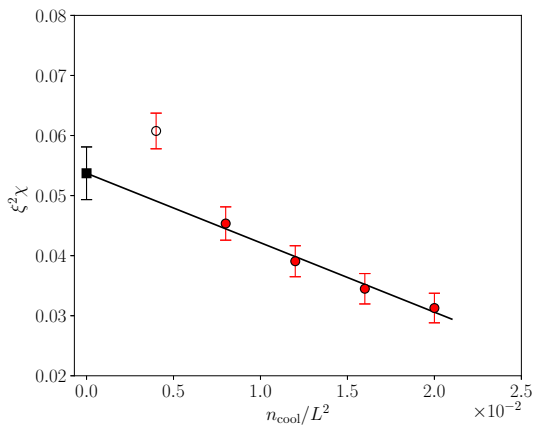}
\caption{Zero-cooling extrapolation of continuum -extrapolated results for $\xi^2 \chi(n_\cool)$ for $N=2$ and $L=50$ obtained for $n_\cool=20,30,40,50$ (full points). Determination for $n_\cool=10$ has been excluded from the fit (empty point). Square full point at $n_\cool=0$ represents our zero cooling extrapolation according to fit function~\eqref{eq:ansatz_zero_cool_extrapolation} without keeping into account correlations among determinations of $\xi^2\chi$ for different values of $n_\cool$. For the final result, see the text and Tab.~\ref{tab:summary_final_results_N_2_L_50_100_200}.}
\label{fig:zero_cool}
\end{figure}

In order to check that all systematics are under control, we repeated this analysis varying the number of cooling steps $n_\cool$, changing the value of $\beta_L^\star$ and narrowing the fit range. While again we observe that the latter two choices do not produce any appreciable change in the obtained result for $a_0 \ne 0$, as any observed variation of this parameter is much smaller compared to its statistical error, we observe a systematic drift of our continuum extrapolations for $\xi^2 \chi$ as $n_\cool$ is increased (see Fig.~\ref{fig:zero_cool}).

Naively, one could think that taking the continuum limit at fixed value of $n_\cool = (r_s/a)^2$ would result in a vanishing smoothing radius. However, since we took $a\to0$ fixing $L=l/a$, the quantity $n_\cool / L^2 = (r_s/l)^2$ is kept constant in our continuum extrapolation and does not disappear from the game.
The fact that $\xi^2 \chi$ decreases increasing $n_{\cool}$ can be easily understood in these terms: $n_{\cool}$ fixes $r_s$ in lattice spacing units, hence results should eventually become independent of $n_{\cool}$ for a theory with no UV divergences. However, because of the divergent small-instanton density and of the fixed ratio between the UV and the IR cut-offs $\sqrt{n_\cool}/L=r_s/l$, the fraction of topological signal which is smoothed away becomes eventually finite and independent of the lattice spacing, but increases as $r_s/l$ increases.

In order to provide the correct final result for $\xi^2 \chi(N=2)$ including the full UV contribution, the correct thing to do is to extrapolate continuum results towards the $n_\cool \to 0$ limit. To do so, we extrapolated our continuum determinations for $\xi^2 \chi$ assuming the following scaling function:
\beq\label{eq:ansatz_zero_cool_extrapolation}
\xi^2\chi\left(\frac{n_\cool}{L^2}\right) = \xi^2\chi\left(\frac{n_\cool}{L^2} = 0\right) + A \frac{n_\cool}{L^2}.
\eeq

This fit function is justified on the basis of the argument explained in Ref.~\cite{Altenkort:2020axj}. Such argument is, strictly-speaking, proven within the gradient flow formalism. However, since it has been shown that performing $n_\cool$ cooling steps is numerically equivalent to flow for a time $\tau_{\mathrm{flow}} = k\, n_\cool$ with $k$ constant (e.g., $\tau_{\mathrm{flow}} = n_\cool/3$ in the $4d$ $\SU(3)$ pure-gauge theory with the Wilson action)~\cite{Bonati:2014tqa}, we expect this argument to also apply in our case.

In the continuum theory, any operator $\mathcal{O}_{\mathrm{smooth}}$ computed on smoothed fields can be expressed in terms of operators computed on the non-smoothed ones by the OPE (Operator Product Expansion) formalism. The leading order contribution is simply given by $\mathcal{O}$ computed on non-smoothed fields (apart from a multiplicative renormalization constant), and higher-order contributions coming from contaminating higher-dimensional operators are suppressed as suitable compensating powers of the amount of smoothing performed. In the case of the topological susceptibility, the relevant operator to be considered is just the topological charge density $q(x)$, since $\chi = \int d^2x \, \braket{q(x)q(0)}$. In this case the renormalization constant appearing in front of the leading order term is just 1 because of the non-renormalizability of the topological charge in the continuum theory, while the next-to-leading order term is suppressed as $n_\cool \propto r_s^2$~\cite{Altenkort:2020axj}. This justifies the ansatz given in Eq.~\eqref{eq:ansatz_zero_cool_extrapolation}.

The result of the best fit of our data with ansatz~\eqref{eq:ansatz_zero_cool_extrapolation} is shown in Fig.~\ref{fig:zero_cool}. A linear 
term in $n_\cool$ nicely describes our data for $n_\cool > 10$ ($\chi^2/\dof = 0.32 / 2$). Including $n_\cool=10$ instead yields a 
much larger $\chi^2/\dof = 7.4 / 3$, thus providing a worse description of our data. Excluding further points (e.g., $n_\cool=20, 30$) gives compatible results within the errors with the one obtained excluding $n_\cool=10$, thus justifying our choice for the fit range. Our final zero cooling extrapolation turns out to be $\xi^2\chi(n_\cool = 0)=0.054(4)$, i.e., clearly different from zero. The latter result has been obtained by performing the continuum extrapolation at fixed $n_\cool$ followed by the $n_\cool/L^2 \to 0$ limit on $\mathcal{O}(1000)$ bootstrap resamplings extracted for each value of $\xi_L$, each one of the same size of the corresponding original dataset.

\subsection{Checking the $L$ dependence and the thermodynamic limit}
\label{subsec:comparison_old_work}

In Sec.~\ref{subsec:multicano_results} we have shown that our results for the topological susceptibility are compatible with the $\log$-divergent continuum limit predicted by semiclassical arguments. Our numerical evidence has been obtained 
on lattices with fixed $L = 50$, i.e., with vanishing volume in the 
continuum limit, and is based on the ansatz, stemming from 
perturbative computations, reported in 
Eqs.~\eqref{eq:chi_small_N_continuum_scaling} and~\eqref{eq:small_N_UV_corrections}. Therefore, as a last step  along our investigation, it is useful to check the 
dependence on $L$ appearing in this ansatz and, moreover, that results are consistent with those obtained  in standard simulations approaching the thermodynamic limit,
i.e., for fixed $l = L a$ and $L \gg \xi_L$, such as those reported
in our previous study in Ref.~\cite{Berni:2020ebn}. 

As already discussed in Sec.~\ref{subsec:strategy}, we have the following prediction (see Eq.~(\ref{eq:chi_small_N_continuum_scaling})):
\beq
\xi^2 \chi = C \log\left(\frac{\rho_{\max}}{\rho_{\min}}\right),
\eeq
where $\rho_{\max} / \rho_{\min}$ are 
the maximum/minimum instanton size that can be observed on 
the given lattice. On a small lattice with fixed $L \ll \xi_L$,
we expect $\rho_{\max} \propto L$, while on a large
lattice with $L \gg \xi_L$ we expect $\rho_{\max}$ to be 
fixed by some physical IR cut-off, hence $\rho_{\max} \propto \xi_L$ in lattice
spacing units. Regarding $\rho_{\min}$, instead, we expect it to be 
proportional to the lattice spacing, with a proportionality 
constant independent of $L$ as long as $L \gg 1$. 
Putting these considerations together, we expect:
\beq
\label{eq:chi_divergence_xi}
\xi^2\chi(\xi_L) &\underset{\xi_L\to\infty}{\sim}& C \log\left(\frac{\xi_L}{\overline{R}}\right), \quad L\gg\xi_L, \\
\label{eq:chi_divergence_L}
\xi^2\chi(L)     &\underset{\xi_L\to\infty}{\sim}& C \log\left(\frac{L}{R}\right), \quad L\ll \xi_L,
\eeq
where $R$ and $\overline{R}$ are two effective parameters which 
are different in the two cases, while the pre-factor
$C$ is expected (and we will actually check) to be the same, since 
it just comes from the (unknown) pre-factor of the instanton density $d_I(\rho) \propto 1/\rho$.

To extract $C$ from finite $L$ results, thus, we need to study the $L$-dependence of the finite continuum limit of $\xi^2 \chi(N=2)$. For this reason, we also performed simulations for $L=100$ and $L=200$. The only difference compared to the $L=50$ investigation discussed above
is that, for these lattices, we do not employ the multicanonical algorithm, 
since the logarithmic UV-divergence is assumed \emph{a priori}, hence 
we do not need extremely precise data to disprove a convergent behavior. In Tab.~\ref{tab:summary_results_N_2_L_100_200} we summarize the parameters of the simulations performed for $L=100$ and $200$.
\begin{table*}[!htb]
\begin{center}
\begin{tabular}{|c|c|c|c|c|c|c|c|c|}
\hline
\multirow{2}{*}{$N$} & \multirow{2}{*}{$\beta_L$} & \multirow{2}{*}{$\xi_L$} & \multicolumn{3}{c|}{$L=100$} & \multicolumn{3}{c|}{$L=200$} \\ \cline{4-9}
&&&&&&&&\\[-1em]
& & & \multicolumn{1}{c|}{$a^2 \chi\cdot 10^{9}$} & \multicolumn{1}{c|}{$\xi^2 \chi\cdot 10^{3}$} & Stat. & \multicolumn{1}{c|}{$a^2 \chi\cdot 10^{9}$} & \multicolumn{1}{c|}{$\xi^2 \chi\cdot 10^{3}$} & Stat. \\ \hline
\multirow{17}{*}{2}
& 1.70 &  0.17991(78) & 3912(10)  & 126.6(1.2) & \multirow{8}{*}{9M} & 5620(22) & 181.9(1.7) & \multirow{8}{*}{2M} \\
& 1.75 & 0.2393(10) & 2149.1(7.9) & 123.0(1.2) &  & 3097(16)   & 177.3(1.8) & \\
& 1.80 & 0.3185(14) & 1193.5(5.9) & 121.1(1.2) &  & 1717(12)   & 174.2(2.0) & \\
& 1.85 & 0.4243(18) & 652.9(4.4)  & 117.5(1.3) &  & 948.5(9.4) & 170.7(2.3) & \\
& 1.90 & 0.5656(25) & 365.9(3.4)  & 117.1(1.5) &  & 525.0(6.9) & 168.0(2.7) & \\
& 1.95 & 0.7545(33) & 198.9(2.5)  & 113.3(1.7) &  & 282.9(5.4) & 161.0(3.4) & \\
& 2.00 & 1.0072(44) & 108.3(1.9)  & 109.9(2.2) &  & 160.4(4.1) & 162.7(4.4) & \\
& 2.05 & 1.3453(58) & 61.0(1.5)   & 110.4(2.8) &  & 88.5(2.3)  & 160.1(4.4) & \\
\cline{6-6}\cline{9-9}
& 2.10 & 1.7980(78) & 33.81(56) & 109.3(2.0) & \multirow{2}{*}{37M} & 48.2(1.2) & 155.8(4.2) & 4M\\
\cline{9-9}
& 2.15  & 2.404(10) & 18.23(39) & 105.4(2.4) & & 27.57(87) & 159.4(5.2) & \multirow{2}{*}{8M} \\
\cline{6-6}
& 2.20  & 3.217(14) & 10.01(21) & 103.6(2.4) & \multirow{7}{*}{75M} & 14.40(43)    &    149.0(4.7) & \\
\cline{9-9}
& 2.25  & 4.307(19) & 5.61(16) & 104.0(3.1) & & 9.28(43) & 172.0(8.2) & \multirow{6}{*}{15M} \\
& 2.30  & 5.768(25)  & 3.08(12)  & 102.5(4.2) & & 4.43(26)  & 147.5(8.6) & \\
& 2.35  & 7.729(34)  & 1.745(93) & 104.3(5.6) & & 2.53(22)  & 151(13)    & \\
& 2.40  & 10.362(45) & 0.955(72) & 102.6(7.8) & & 1.30(12)  & 139(13)    & \\
& 2.45  & 13.897(60) & 0.542(48) & 104.6(9.3) & & 0.767(86) & 148(17)    & \\
& 2.50  & 18.645(81) & 0.355(47) & 123(16)    & & 0.458(74) & 159(26)    & \\
\hline
\end{tabular}
\end{center}
\caption{Summary of simulation parameters and results obtained for $N=2$ and $L=100,200$. The correlation length $\xi_L$ is computed according to Eq.~\eqref{eq:xi_2loop} with $\beta_L^\star = 1.70$. Reported values of $\chi$ are computed after $n_\cool = 50$ cooling steps. Statistics is expressed in millions (M) and measures are taken every 10 standard MC steps ($=$ 4 OR + 1 HB lattice updating sweeps).}
\label{tab:summary_results_N_2_L_100_200}
\end{table*}

\begin{figure}[!htb]
\centering
\includegraphics[scale=0.43]{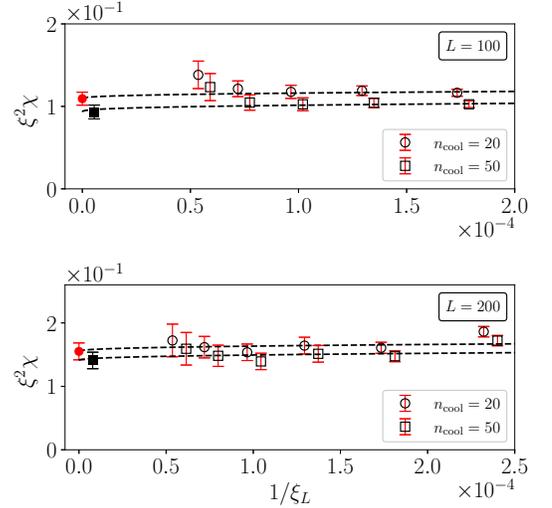}
\caption{Extrapolation towards the continuum limit of $\xi^2 \chi$ for $N=2$ and $L=100, 200$ for $n_{\cool}=20, 50$. Dashed lines represent best fits obtained using fit function~\eqref{eq:ansatz_cont_limit}. The full point in $1/\xi_L = 0$ represents the non-vanishing continuum-extrapolated determination.}
\label{fig:fit_continuum_limit_N_2_L_100_200}
\end{figure}

The computation of $\xi^2\chi$ for $L=100,200$ has been done following the same lines of Sec.~\ref{subsec:multicano_results}. First, we extrapolate our results towards the continuum limit at fixed value of $n_\cool$. Continuum extrapolations at fixed $n_\cool = 20,50$ for $L=100,200$ are shown in Fig.~\ref{fig:fit_continuum_limit_N_2_L_100_200}. As a further consistency check, we also verified that the free exponent $c$ appearing in the fit function in Eq.~\eqref{eq:ansatz_cont_limit} was compatible within the errors in all cases, cf.~Tab.~\ref{tab:summary_final_results_N_2_L_50_100_200}.

\begin{figure}[!htb]
\centering
\includegraphics[scale=0.43]{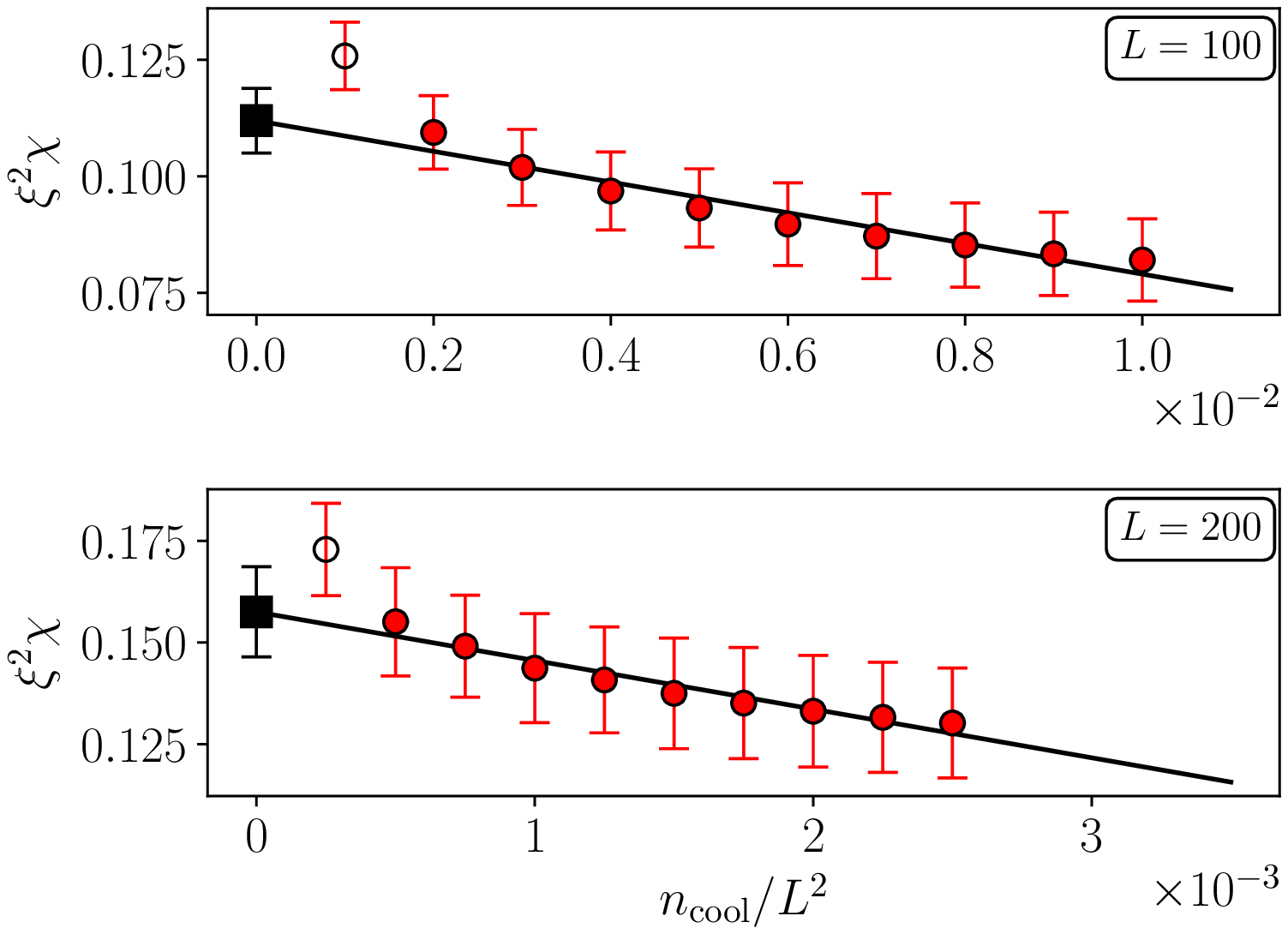}
\caption{Zero-cooling extrapolation of continuum-extrapolated results for $\xi^2 \chi(n_\cool)$ for $N=2$ and $L=100,200$ obtained for $n_\cool=20,30,\dots,100$ (full points). Determinations for $n_\cool=10$ have been excluded from the fit (empty point). Square full point at $n_\cool=0$ represents our zero cooling extrapolation according to fit function~\eqref{eq:ansatz_zero_cool_extrapolation} without keeping into account correlations among determinations of $\xi^2\chi$ for different values of $n_\cool$. For the final result, see the text and Tab.~\ref{tab:summary_final_results_N_2_L_50_100_200}.}
\label{fig:zero_cool_L_100_200}
\end{figure}

Then, we extrapolate such continuum determinations towards the zero-cooling limit. Again, our results for $L=100,200$ are nicely described by a linear function in $n_\cool/L^2$, cf.~Fig.~\ref{fig:zero_cool_L_100_200}. Our final results for $\xi^2\chi(n_\cool/L^2=0)$, for $L=50, 100$ and $200$ are collected in Tab.~\ref{tab:summary_final_results_N_2_L_50_100_200}.

\begin{table}[!htb]
\begin{center}
\begin{tabular}{|c|c|c|}
\hline
$L$ & $\xi^2\chi(n_\cool/L^2 = 0)$	& exponent $c$ \\
\hline
50  & 0.054(4) & 0.20(2)  \\
100 & 0.109(9) & 0.35(14) \\
200 & 0.15(2)  & 0.39(20)  \\
\hline
\end{tabular}
\end{center}
\caption{Double extrapolated results for $\xi^2 \chi(L)$ ($1/\xi_L\to 0$ followed by $n_\cool/L^2\to 0$) and determinations of the exponent $c$ appearing in fit function~\eqref{eq:ansatz_cont_limit}.}
\label{tab:summary_final_results_N_2_L_50_100_200}
\end{table}

\begin{figure}[!htb]
\centering
\includegraphics[scale=0.46]{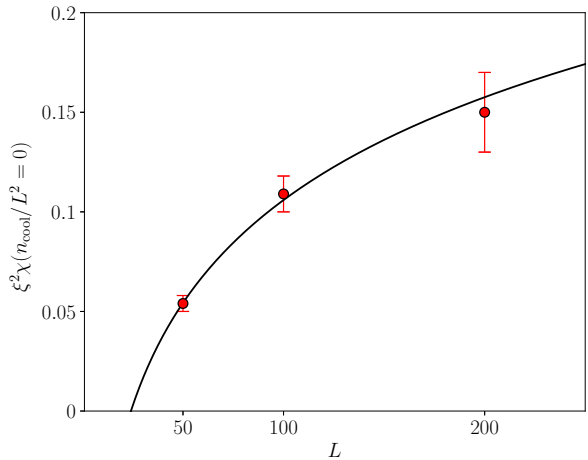}
\caption{Best fit of $\xi^2\chi(n_\cool/L^2=0)$ as a function of $L$ according to Eq.~\eqref{eq:chi_divergence_L}. Best fit gives $\tilde{\chi}^2/\dof=0.23/1$}.
\label{fig:chi_zero_nc_vs_L}
\end{figure}

Finally, we performed a best fit of our results for $\xi^2\chi(n_\cool/L^2=0)$ as a function of the fixed lattice size $L$ according to Eq.~\eqref{eq:chi_divergence_L} to determine the pre-factor $C$.

Our data are very-well described by a log-divergent function of the lattice size $L$, as shown in Fig.~\ref{fig:chi_zero_nc_vs_L}, and we obtain:
\beq
\label{eq:C_small_L}
C &=& 0.074(11),\\
\label{eq:R_small_L}
R &=& 24(3).
\eeq
It is now interesting to compare these results with those of Ref.~\cite{Berni:2020ebn}, obtained in the thermodynamic limit $L/\xi_L\gg 1$.

Extrapolating the results for $L/\xi_L \gtrsim 12$ reported in that work towards the continuum limit, and according to the divergent fit function in Eq.~\eqref{eq:chi_divergence_xi} plus $O(\xi_L^{-2})$ corrections, we obtain
\beq
\label{eq:C_large_L}
C &=& 0.074(2),\\
\label{eq:R_large_L}
\overline{R} &=& 4.7(3).
\eeq
As expected, while the constants $R$ and $\overline{R}$ are different, the pre-factors $C$ of the logarithms turn out to be in perfect agreement among each other.

\begin{figure}
\centering
\includegraphics[scale=0.39]{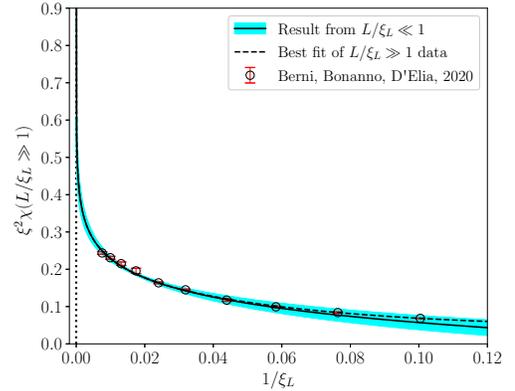}
\caption{Results for $\xi^2\chi$ reported in Ref.~\cite{Berni:2020ebn} for $L/\xi_L\gg 1$ and $n_\cool = 20$. Dashed line represents best fit of these data according to fit function $f(x) = -C \log(\overline{R} x) + C_2 x^2$ where $x = 1/\xi_L$ and the result for $C$ is reported in Eq.~\eqref{eq:C_large_L}. Solid line represents the curve $g(x) = -C \log(R x) + \log(R/\overline{R})$, where $x = 1/\xi_L$ and $C$ is reported in Eq.~\eqref{eq:C_small_L}. The shadowed area represents the error band on $g(x)$.}
\label{fig:divergence_xi_old_new_comp}
\end{figure}

In Fig.~\ref{fig:divergence_xi_old_new_comp} we show the $L/\xi_L\gg 1$ determinations for $\xi^2\chi$ of Ref.~\cite{Berni:2020ebn} along with their best fit according to Eq.~\eqref{eq:chi_divergence_xi} plus $O(\xi_L^{-2})$ corrections. On top of these, we plot the curve $C\log(\xi_L/R) + \log(R/\overline{R})$, using the value of $C$ in Eq.~\eqref{eq:C_small_L}, i.e., coming from the logarithmic best fit of the fixed-$L$ results obtained in this work, reported in Tab.~\ref{tab:summary_final_results_N_2_L_50_100_200}. The two curves collapse on top of each other.

In conclusion, the comparison carried out in this subsection provides 
solid numerical evidence that results obtained by fixed $L$ simulations 
contain information which is consistent, as for the UV-behavior of the 
topological susceptibility, with what would be obtained in 
the thermodynamic infinite volume limit.

\section{Conclusions}\label{sec:final}

The purpose of the present study was that of providing numerical evidence for the 
predicted divergent behavior in the continuum limit of the topological susceptibility
of the $\CP^1$ model. The same problem has been considered by several past studies, the novelty of the present investigation is to approach the continuum limit at fixed volume in 
dimensionless lattice units: this maps a logarithmically divergent behavior, 
which can be barely distinguishable from a badly convergent behavior 
over a wide range of lattice spacings, into a convergent behavior with a 
non-vanishing continuum limit, which is more amenable to be checked numerically
with a well definite conclusion.

After checking that this method reproduces the results obtained with standard strategies 
for the $\CP^2$ and the $\CP^3$ theories, we applied it to our target model,
implementing at the same time a multicanonical algorithm in order solve the problem of 
rare fluctuations of the topological charge on asymptotically small lattices. The use of the multicanonical algorithm revealed essential, since it reduced the computational effort by up to two order of magnitudes for the smallest explored lattice spacings.

Our results show that the continuum limit of the topological susceptibility 
of the $\CP^1$ model obtained at fixed $L$, and after extrapolation 
to zero cooling steps, is indeed non-vanishing, as predicted by semi-classical
computations. Moreover, 
repeating the same computation for different values of $L$,
we observe that the obtained non-vanishing determinations of $\xi^2 \chi$
grow proportionally to $\log L$, with a prefactor consistent with previous lattice
results: 
that provides evidence that our investigation at fixed $L$ is perfectly consistent
with what would be obtained in 
the thermodynamic infinite volume limit; however
it permits, at the same time, to definitely disprove 
the possibility of a convergent behavior for $\chi$.

\acknowledgments

The authors thank C.~Bonati and P.~Rossi for useful discussions. CB acknowledges the support of the Italian Ministry of Education, University and Research under the project PRIN 2017E44HRF, ``Low dimensional quantum systems: theory, experiments and simulations''. Numerical simulations have been performed on the \texttt{MARCONI} machine at CINECA, based on the agreement between INFN and CINECA (under projects INF21\_npqcd and INF22\_npqcd).

\appendix

\section{Asymptotic scaling check}
\label{app:asymptotic_scaling}

To check if our assumption of being in the asymptotic scaling region is correct, we consider the quantity $M/\Lambda^{\mathrm{(Sym)}}_L \equiv [\xi_L f(\beta_L)]^{-1}$, which is expected to be constant plus $O(1/\beta_L)$ corrections for $\beta_L\to\infty$. For $N=2$, the exact value of $M$ in the continuum is known and can be expressed in terms of the dynamically-generated scale of the Symanzik theory $\Lambda_L^{\mathrm{(Sym)}}$ by combining results of Refs.~\cite{Hasenfratz:1990zz,Campostrini:1992ar}:
\beq\label{eq:exact_result}
\frac{M}{\displaystyle\Lambda_L^{\mathrm{(Sym)}}}(N=2) \bigg\vert_{\mathrm{exact}} \simeq 21.7.
\eeq

To test asymptotic scaling for $\CP^{1}$, $\CP^{2}$ and $\CP^{3}$ models, we consider results for $\xi_L$ as a function of $\beta_L$ of~\cite{Berni:2020ebn}. Furthermore, we also added higher-$\xi_L$ data to this analysis, which are reported in Tab.~\ref{tab:new_data}. 

\begin{table}[!htb]
	\begin{center}
		\begin{tabular}{|c|c|c|c|c|}
			\hline
			$N$ & $\beta_L$ & $L$ & $\xi_L$ & $L/\xi_L$ \\
			\hline
			\multirow{8}{*}{2} & \multirow{8}{*}{1.70}
			& 360      & 141.93(26) & 2.5      \\
			& & 500      & 162.13(63) & 3        \\
			& & 600      & 170.17(66) & 3.5      \\
			& & 700      & 174(1)     & 4        \\
			& & 800      & 177(1)     & 4.5      \\
			& & 1024     & 179.81(87) & 5.7      \\
			& & 1450     & 181(2)     & 8        \\
			& & $\infty$ & 179.91(78) & $\infty$ \\
			\hline
			\multirow{2}{*}{3}
			& 1.32  & 562  & 44.75(29) & 12.5 \\
			& 1.455 & 1250 & 98.19(53) & 12.4 \\
			\hline
			\multirow{3}{*}{4}
			& 1.20  & 436  & 34.03(18)  & 12.7 \\ 
			& 1.30 & 766   &  61.76(34) & 12.5 \\
			& 1.35 & 1030  & 82.62(70)  & 12.5 \\
			\hline
		\end{tabular}
	\end{center}
	\caption{Simulation summary of the additional runs performed to check asymptotic scaling for $N=2,3$ and $4$.}
	\label{tab:new_data}
\end{table}

\begin{figure}[!htb]
\centering
\includegraphics[scale=0.3]{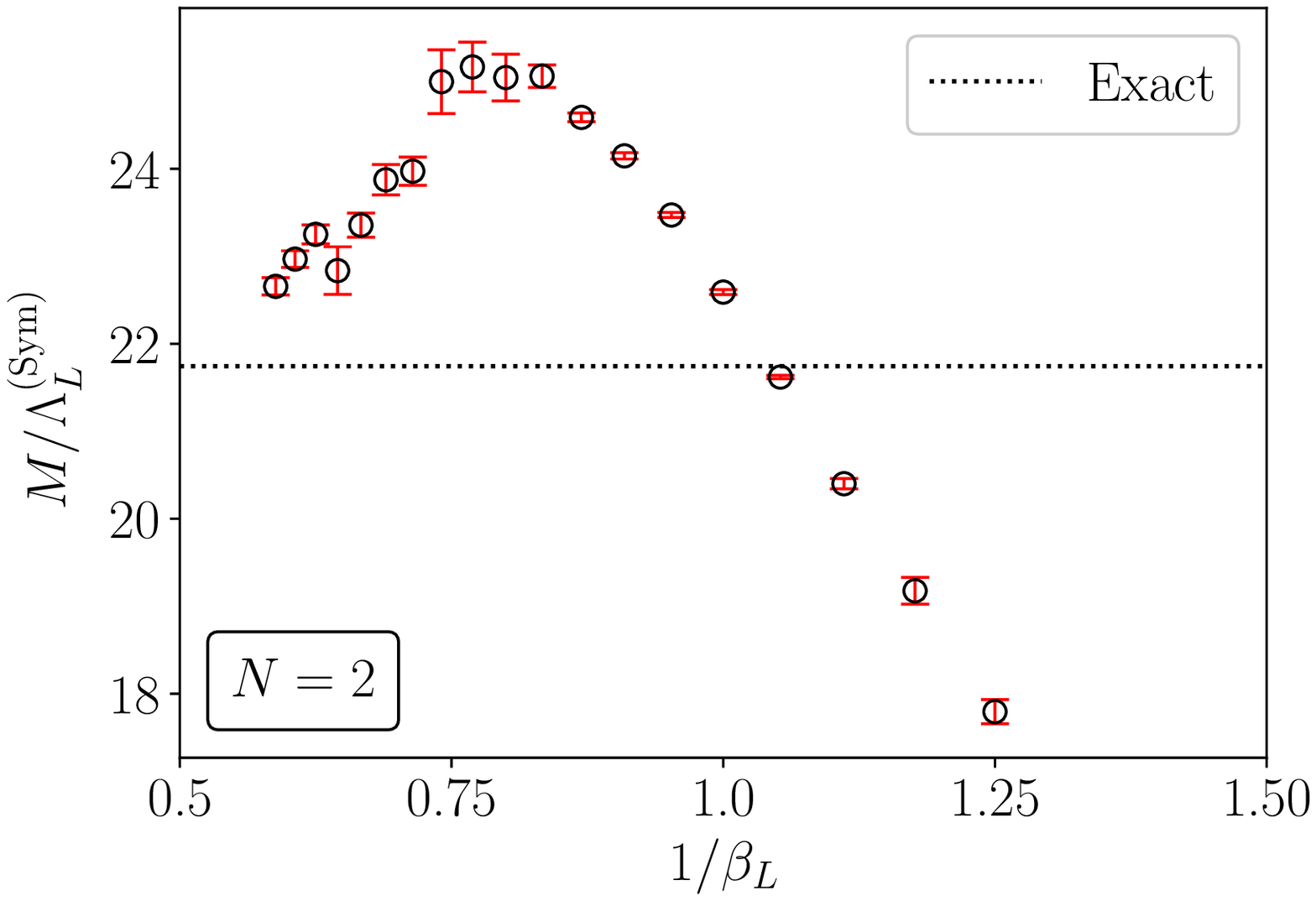}
\includegraphics[scale=0.3]{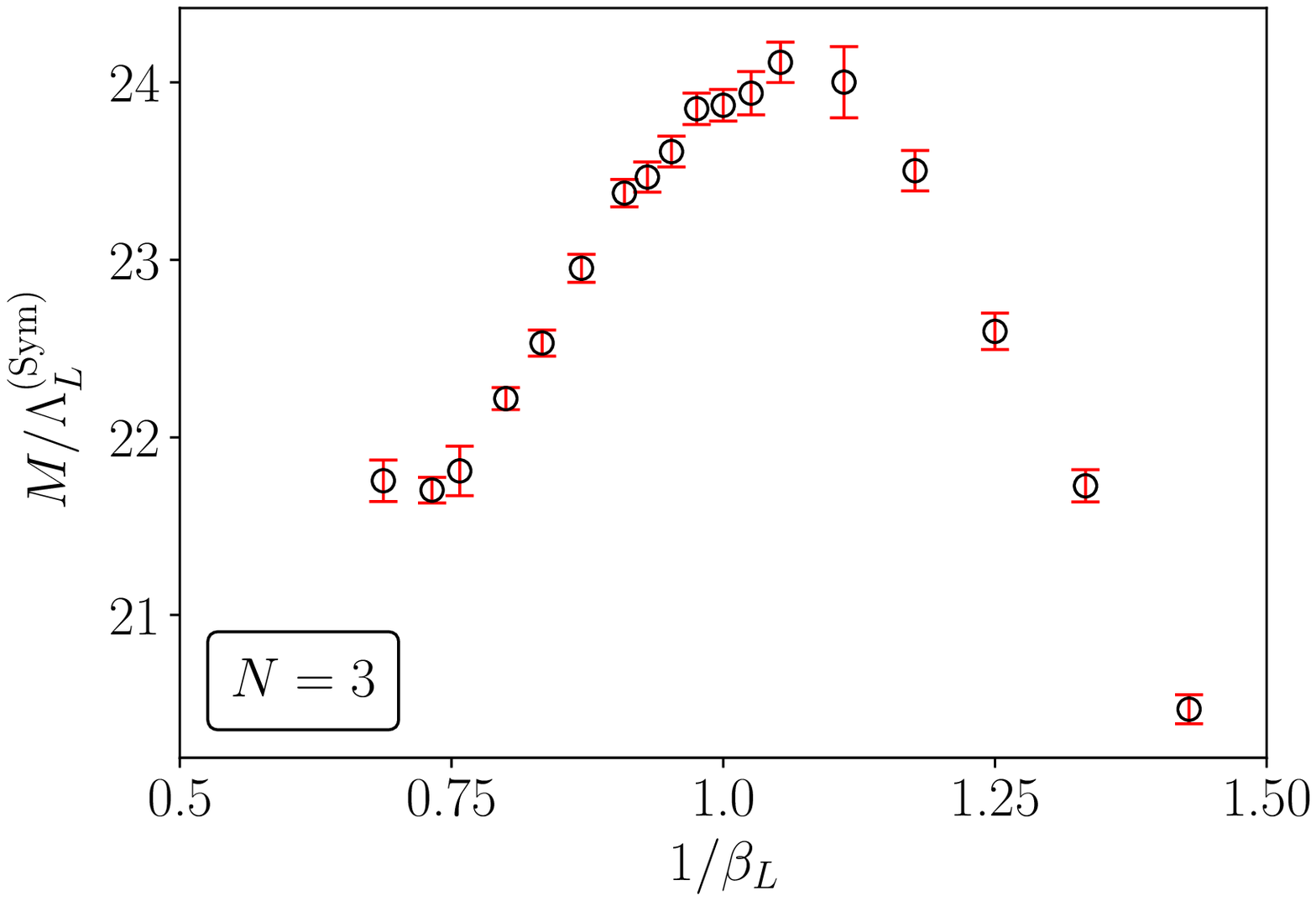}
\includegraphics[scale=0.3]{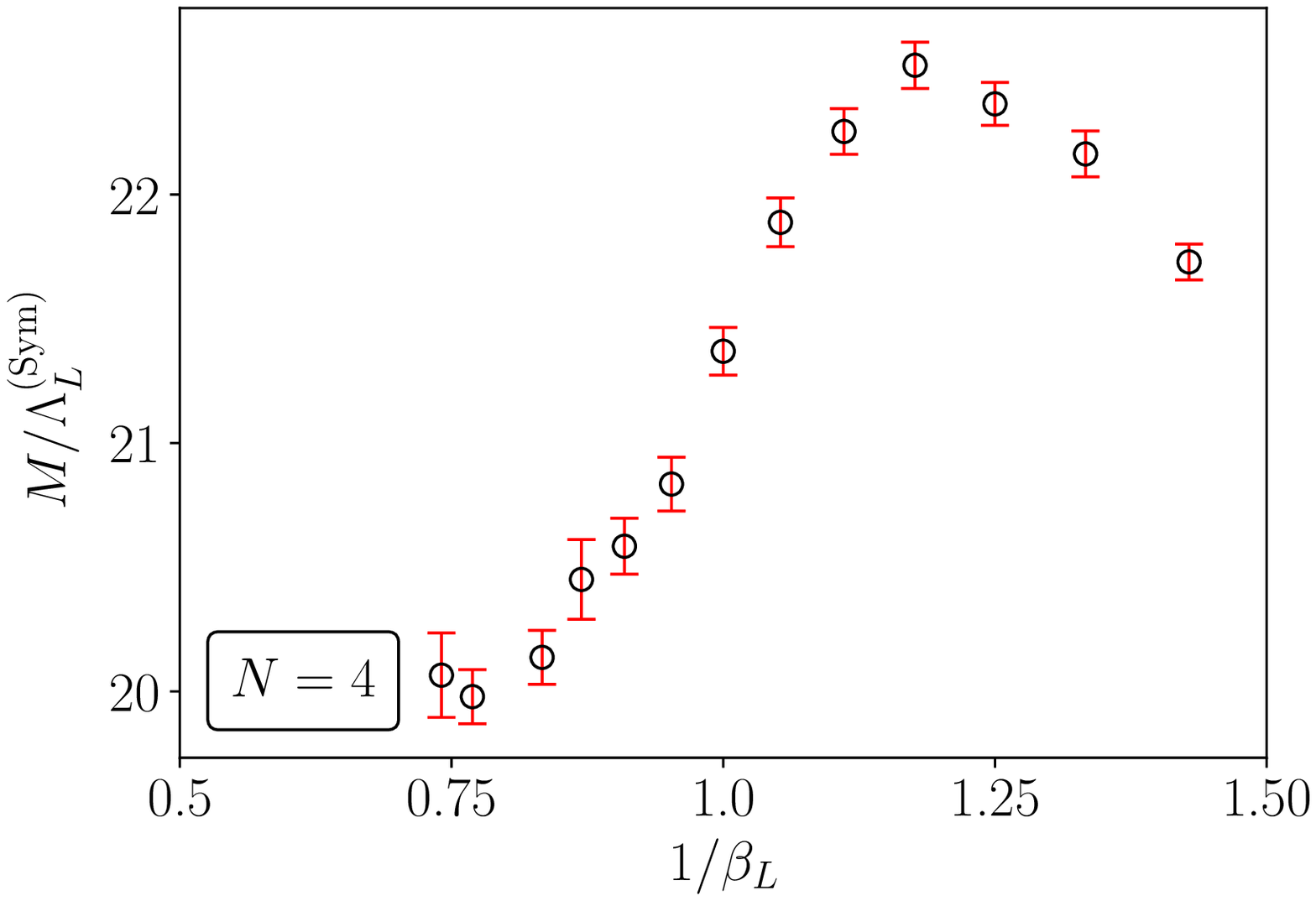}		
\caption{Check of the asymptotic scaling of $\xi_L$ for the $\CP^1$, $\CP^2$ and $\CP^3$ models. The figures show the behavior of $M/\Lambda^{\mathrm{(Sym)}}_L = [\xi_L f(\beta_L)]^{-1}$ as a function of the inverse coupling $1/\beta_L$. For $N=2$, the dotted line displays the exact analytic result for the continuum limit of $M/\Lambda_L^{\mathrm{(Sym)}}(N=2)$ in Eq.~\eqref{eq:exact_result}.}
\label{fig:asymptotic_scaling_N}
\end{figure}

For $N=3$ and $4$, we chose the lattice size requiring that $L/\xi_L \gtrsim 12$, which is enough to ensure that finite size effects are well under control. For $N=2$, instead, we computed $\xi_L$ for several lattice sizes and extrapolated it towards the thermodynamic limit by fitting its $L$-dependence according to:
\beq
\label{eq:finite_size_scaling}
\xi_L(L) = \xi_L^{(\infty)} (1 - a \, e^{-b \, L/\xi_L}),
\eeq
where $\xi_L^{(\infty)}$ is the desired quantity and $a$ and $b$ are additional fit parameters. In Figs.~\ref{fig:asymptotic_scaling_N} we display the quantity $M/\Lambda^{\mathrm{(Sym)}}_L = [\xi_L f(\beta_L)]^{-1}$ as a function of $1/\beta_L$ for, respectively, the $\CP^1$, $\CP^2$ and $\CP^3$ models. 

For $N=4$ and 3 the quantity $M/\Lambda^{\mathrm{(Sym)}}_L$ reaches a plateau asymptotically. Thus, we choose $\beta_L^\star(N=4) = 1.35$ and $\beta_L^\star(N=3)=1.455$ to fix $\xi_L$ via Eq.~\eqref{eq:xi_2loop}. For $N=2$, despite the wider range of $1/\beta_L$ explored, we observe a slower approach to the asymptotic scaling regime probably due to larger $O(1/\beta_L)$ corrections in this case. Nonetheless, we observe that the obtained results for $\xi_L$ using Eq.~\eqref{eq:xi_2loop} do not show an appreciable dependence on the choice of $\beta_L^\star$, showing that our procedure to fix the scale is solid even in this case. As an example, for $N=2$ and $\beta_L=3.00$ we have $\xi_L = 354.7(1.5)$ if $\beta_L^\star = 1.65$ and $\xi_L = 359.6(1.6)$ if $\beta_L^\star = 1.70$, i.e., the two determinations agree within $\sim2.2$ standard deviations. Therefore, we choose $\beta_L^\star(N=2) = 1.70$ to fix the scale in this case.

\FloatBarrier

\section{Multicanonical algorithm details}
\label{app:details_multicano_algorithm}

The topological bias potential was chosen according to the same functional form adopted in Ref.~\cite{Bonati:2018blm}:
\beq 
\label{eq:topo_potential}
V_{\topo}(x)  =
\begin{cases}
\displaystyle- \sqrt{(B x)^2 + C}, & \mathrm{if} \ \vert x \vert \le Q_{\max}, \\ 
\\
\displaystyle- \sqrt{(B Q_{\max})^2 + C}, & \mathrm{if} \ \vert x \vert > Q_{\max}. \\ 
\end{cases}
\eeq

\begin{figure}[!b]
	\centering
	\includegraphics[scale=0.35]{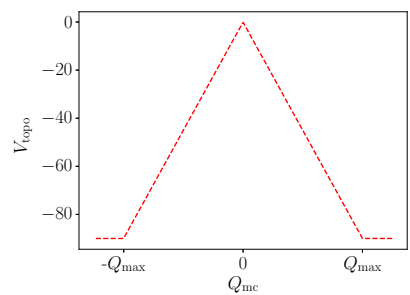}
	\caption{Illustrative example of the bias potential $V_{\topo}(Q_{mc})$ in Eq.~\eqref{eq:topo_potential} with $B=7.5$, $C=0.05$ and with a cut at $Q_{\max}=12$.}
	\label{fig:topo_potential}
\end{figure}

Here, $B$, $C$ and $Q_{\max}$ are free parameters that can be calibrated through short preliminary runs to improve the performances of the multicanonic algorithm. The employed values of $B$ varied between $\sim 5$ and $\sim 10$ for $\beta_L \in [2.2, 3]$, while the choice of $C$ and $Q_{\max}$ turned out to be not critical, thus we used $C=0.05$ and $Q_{\max}=12$ for all $\beta_L$. An illustrative example of the functional form in Eq.~\eqref{eq:topo_potential} is shown in Fig.~\ref{fig:topo_potential}.

Our implementation of the multicanonic algorithm follows the lines of Ref.~\cite{Jahn:2018dke}. First, we generate a candidate new lattice configuration by performing a standard updating step and ignoring the $Q$-dependent bias potential. Then, we accept the updated configuration by performing a standard Metropolis test:
\beqnn
p &= \min{\left\{1, \exp{(- \Delta V_{\topo})}\right\}},
\eeqnn
where
\beqnn
\Delta V_{\topo} &\equiv V_{\topo}\left(Q_\mc^{(\mathrm{new})}\right) - V_{\topo}\left(Q_\mc^{(\mathrm{old})}\right)
\eeqnn
is the variation of the bias potential before and after the update. After running some preliminary simulations, we found that the optimal implementation to have higher Metropolis acceptances was to perform the Metropolis test after each single-link update $U_\mu(x)$, instead of performing it after a whole standard MC step (i.e., after 5 sweeps of the whole lattice). Moreover, we also found that proposing single-site/single-link updates stochastically was more effective to obtain higher Metropolis acceptances than performing lattice sweeps.

It is easy to verify that, for any starting updating step, our choice respects \emph{detailed balance} for the original distribution. Moreover, when considering the path-integral probability distribution obtained with the modified action in Eq.~\eqref{eq:multicano_action}, our multicanonical updating step with the addition of the Metropolis test respects detailed balance too.

Finally, regarding the topological charge discretization $Q_\mc$, our choice is $Q_\mc = Q_U$, i.e., the geometric definition in Eq.~\eqref{eq:topo_charge_geo} computed without performing any cooling step. This choice allows to avoid the full computation of $Q_\mc$ (necessary to compute the Metropolis probability) every time an update of a link variable $U_\mu(x)$ is proposed, as with this choice one can directly compute $\Delta Q_\mc = Q_\mc^{(\mathrm{new})} - Q_\mc^{(\mathrm{old})}$ in terms of the new link and its relative staples.

With this setup, we obtained mean Metropolis acceptances larger than $90\%$, and we found that a multicanonical MC step required a $\approx 85\%$ larger numerical effort compared to a standard MC step. After taking into account such overhead, we found that the multicanonic algorithm allowed to gain up to two orders of magnitude in terms of computational power compared to the standard algorithm when $\braket{Q^2} \ll 1$.

\end{document}